\newcommand{\Ha}{H$\alpha$}
\newcommand{\Hb}{H$\beta$}

\newcommand{\OII}{O~{\sc ii}}
\newcommand{\OI}{O~{\sc i}}

\newcommand{\NII}{[N~{\sc ii}]}

\newcommand{\SII}{S~{\sc ii}}

\newcommand{\OIII}{[O~{\sc iii}]}

\newcommand{\flux}{\hbox{erg~cm$^{-2}$~s$^{-1}$}}
\newcommand{\lumin}{{erg~s$^{-1}$}}

\newcommand{\msun}{\hbox{${M}_{\odot}$}}

\newcommand{\simgt}{\lower 2pt \hbox{$\, \buildrel {\scriptstyle >}\over {\scriptstyle\sim}\,$}}
\newcommand{\simlt}{\lower 2pt \hbox{$\, \buildrel {\scriptstyle <}\over {\scriptstyle\sim}\,$}}

\newcommand{\xmm}{{\emph{XMM-Newton}}}
\newcommand{\chandra}{{\emph{Chandra}}}

\newcommand{\rosat}{{\emph{ROSAT}}}
\newcommand{\einstein}{{\emph{Einstein}}}

\newcommand{\hst}{{\emph{HST}}}

\newcommand{\sirtf}{{\emph{SIRTF}}}

\documentclass{aastex}
\usepackage{emulateapj5,times,mathptm}
\usepackage{natbib,psfig}

\begin{document}


\title{The Chandra Deep Field-North Survey. XV. \\
  Optically Bright, X-ray Faint 
Sources$^{1}$ }


\author{A.E.~Hornschemeier,$^{2,3}$ ~
F.E.~Bauer,$^2$
D.M.~Alexander,$^2$
W.N.~Brandt,$^2$
W.L.W.~Sargent,$^{4}$
M.W.~Bautz,$^5$
C.~Conselice,$^{4}$
G.P.~Garmire,$^2$
D.P.~Schneider,$^2$
and G.~Wilson$^6$
}

\footnotetext[1]{Based on
observations obtained at the W. M. Keck Observatory which is operated jointly
by the California Institute of Technology and the University of California.  Based  
on observations obtained by the Hobby-Eberly
Telescope which is a joint project of The University of Texas at Austin,
The Pennsylvania State University, Stanford University,
Ludwig-Maximillians-Universit\"at M\"unchen, and Georg-August-Universit\"at
G\"ottingen. Based on observations obtained at the Mayall~4~m telescope,
Kitt Peak National Observatory, 
National Optical Astronomy Observatory, which is operated by the 
Association of Universities for Research in Astronomy, Inc. (AURA) 
under cooperative agreement with the National Science Foundation. }

\footnotetext[2]{Department of Astronomy \& Astrophysics, 525 Davey Laboratory, 
The Pennsylvania State University, University Park, PA 16802}

\footnotetext[3]{Chandra Fellow, Johns Hopkins University, 3400 N. Charles Street,
Baltimore, MD 21218}
\footnotetext[4]{Palomar Observatory, California Institute of Technology, 
Pasadena, CA 91125}

\footnotetext[5]{Massachusetts Institute of Technology, Center for Space Research, 
70 Vassar Street, Building 37, Cambridge, MA 02139}

\footnotetext[6]{Brown University, Physics Department, 182 Hope Street, Providence, RI 02912
and SIRTF Science Center,California Institute of Technology,Mail Code 220-6,1200 East California Boulevard,Pasadena, CA 91125 }


\begin{abstract}
        We have analyzed optically bright, 
X-ray faint [OBXF; i.e., $\log{({{f_{\rm X}}\over{f_{\rm R}}})} \simlt -2$]
 sources identified in an 178.9~arcmin$^2$ area having high 
exposure ($> 1500$~ks)  within the \chandra\ Deep Field-North 
(\hbox{CDF-N}) 2~Ms survey. We find 43 
OBXF sources in this area, comprising 
$\approx 15$\% of the X-ray sources above a 0.5--2~keV flux of 
 $\approx 2.3\times10^{-17}$~erg~cm$^{-2}$~s$^{-1}$.   
We present spectroscopic identifications for 42 of the OBXF 
sources and optical spectra for 25, including 5 previously
unpublished redshifts.  Deep optical imaging data (either \hst\
or ground-based) are presented for all the OBXF sources; we
measure the optical morphologies of the 20 galaxies having \hst\ imaging data.
The OBXF population consists mainly of normal and starburst 
galaxies detected out to cosmologically significant distances
(i.e., to a median redshift of $z=0.297$ and a full redshift
 range $z=0.06$--0.845).  This is notable since these distances 
equate to look-back times of up to $\approx8$~Gyr; we are thus provided
with a window on the X-ray emission from galaxies at redshifts much closer to
the cosmic star formation peak than was possible prior to \chandra. 

The X-ray luminosity distribution of OBXF sources extends 
to higher luminosity than does that of ``normal" galaxies 
indicating that a significant fraction are likely dominated by low-luminosity AGN 
(LLAGN) or vigorous star formation. The lowest redshift galaxies 
($z\approx0.06$--0.2) have very low X-ray-to-optical flux ratios
[i.e., $\log{({{f_{\rm X}}\over{f_{\rm R}}})} \simlt -3$] which 
are consistent with those of normal galaxies in the local Universe.
By combining the detected X-ray counts, 
we find the average OBXF X-ray spectrum to be consistent
with a $\Gamma\approx2.0$ power law. 
The 0.5--2~keV $\log N$--$\log S$ for the OBXF galaxies is much steeper 
($\alpha\approx-1.7$) than for the general X-ray source population.
Indeed, the number of OBXF sources
has doubled between the 1~Ms and 2~Ms survey,
rising sharply in numbers at faint fluxes.
The extragalactic OBXF sources are found to 
contribute $\approx1$--2\% of the soft extragalactic X-ray background.

We report on the discovery of five candidate off-nuclear 
ultraluminous X-ray sources ($L_{\rm X} \simgt 10^{39}$~\lumin) with $z\approx0.1$--0.2
within the OBXF population.  These sources are ``ultraluminous" 
in that they are typically more X-ray luminous than e.g., Eddington-limited 
accretion onto stellar-mass black holes; these sources are found 
to dominate the X-ray emission of their host galaxies.  Included among  
the ULX sources is the remarkable object CXOHDFN~J123721.6+621246, which is located 
along the arm of a grand-design spiral galaxy at $z=0.106$.  This 
object exhibits strong signs of X-ray variability and is likely 
an X-ray binary system.

\end{abstract}




\section{Introduction \label{intro}}

Recent deep X-ray surveys have revealed a population of 
optically bright, X-ray faint sources \citep[e.g.,][hereafter Paper~II; 
Tozzi et~al. 2001]{Horn01}.  The low X-ray-to-optical flux ratios 
[$\log{({{f_{\rm X}}\over{f_{\rm R}}})} \simlt -1$] are
notable because the bulk of the X-ray background is
produced by AGN typically having $-1 < \log{({{f_{\rm X}}\over{f_{\rm R}}})} < +1$.
These sources arise mainly at very faint X-ray fluxes 
\citep[$\simlt 1 \times 10^{-15}$~erg~cm$^{-2}$~s$^{-1}$, 0.5--2.0~keV; 
e.g., Paper II; ][]{Tozzi01} and usually lie at $z \simlt 1$.
Their correspondingly low X-ray luminosities ($L_{\rm X} \simlt 10^{41}$\lumin) 
along with their optically normal spectra indicate that their high-energy 
emission is not obviously dominated by luminous active galactic 
nuclei (AGN).  These sources thus appear to be the distant analogs of
``normal" galaxies in the local Universe.

Previous studies of this population (at shallower depths) have 
necessarily concentrated on intermediate X-ray-to-optical flux ratios [e.g., 
$-2 \simlt \log{({{f_{\rm X}}\over{f_{\rm R}}})} \simlt -1$]
 where the numbers of sources were sufficient for significant astrophysical
constraints to be placed.  
However, more quiescent galaxies (such as our own Milky Way) are expected
to arise at even lower X-ray-to-optical flux ratios [i.e., 
$\log{({{f_{\rm X}}\over{f_{\rm R}}})} \simlt -2$].
Studies at these extremely low X-ray-to-optical flux ratios 
were hardly possible before reaching a depth of 2~Ms 
with $Chandra$ (the number of such sources {\it doubled} between the 1 Ms and 2 Ms samples).
With 2~Ms of $Chandra$ exposure, we find these sources to comprise a significant 
fraction of the most X-ray faint sources
($\approx 22$\% of sources having 
$\approx10^{-17} < f_{\rm X} < 10^{-16}$~erg~cm$^{-2}$~s$^{-1}$, 0.5--2~keV).

Optically bright, X-ray faint (OBXF) objects offer a window 
into the high-energy processes in galaxies at cosmologically 
significant look-back times, allowing the study of 
accretion onto stellar-mass compact objects, 
supernova remnants, low-level accretion onto supermassive 
black holes, and hot gas in the
interstellar media of large early-type galaxies 
over timescales of billions of years.   
The majority of the X-ray sources in the Universe are expected
to be these more quiescent galaxies, while the luminous AGN which dominate
the flux of the X-ray background are a minority.
This study is the distant analog to X-ray surveys of galaxies
in the local Universe
\citep*[e.g.,][]{David92,Read97,Shapley01}.

As mentioned, there has been some work carried out using deep X-ray surveys
to examine sources with
lower X-ray-to-optical flux ratios [$\log{({{f_{\rm X}}\over{f_{\rm R}}})} \simlt -1$]. 
At intermediate values of the X-ray-to-optical flux ratio
[$-2 \simlt \log{({{f_{\rm X}}\over{f_{\rm R}}})} \simlt -1$], 
the majority of X-ray sources has been shown to be considerably
more ``active" than nearby normal galaxies and appear to be
the \hbox{$z=0.4$--1.3} analog of nearby luminous infrared galaxies
(Alexander et al. 2002, hereafter Paper~XI; Bauer et al. 2002, hereafter Paper~XII).
   There have also been statistical
studies of {\it optically selected} galaxies which used X-ray stacking techniques
to probe X-ray emission below the individual point source detection threshold
of, e.g., 1~Ms surveys.  These studies \citep[e.g.,][hereafter Paper~VIII]{Hornstacking}
have probed even lower X-ray-to-optical flux ratios 
[$\log{({{f_{\rm X}}\over{f_{\rm R}}})} \approx -2.8$]
and demonstrated that the 
large population of quiescent field spiral galaxies is much less X-ray
luminous than the typical X-ray detected galaxies 
1~Ms \chandra\ surveys \citep[][hereafter Paper~VIII]{Hornstacking}.

\nocite{davoXI,BauerXII}

\begin{figure}[t!]
\figurenum{1}
\centerline{\includegraphics[scale=0.85,angle=0]{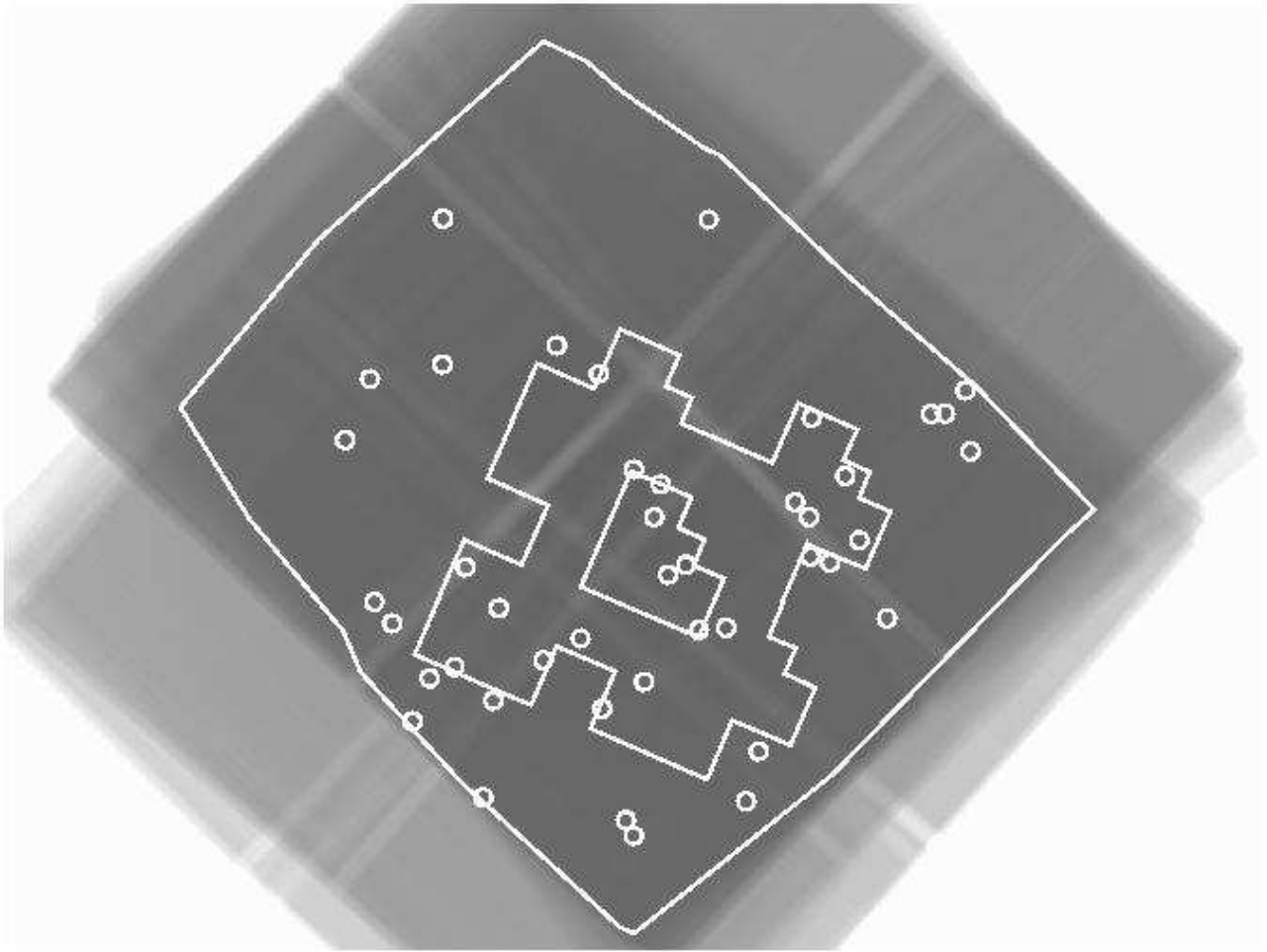}}
\caption[OBXF Exposure Map]{
Full-band exposure map for the 2~Ms CDF-N data set as calculated 
in Paper~XIII. The dimensions of the map are $19^{\prime}\times18^{\prime}$.
   The large polygon indicates the High-Exposure Area 
(HEA) where the typical effective exposure time (excluding
the gaps between the ACIS CCDs) is $> 1.5$~Ms.  
The large, complicated polygon marks the boundary of the \hst\ 
Flanking-Fields coverage, and the small polygon indicates the \hbox{HDF-N}.
  The light grooves running through the exposure 
map are the gaps between the CCDs in the ACIS detector.  The 
circles mark the optically bright, X-ray faint (OBXF) sources;
 each has a radius of 10$^{\prime\prime}$.  There are fewer sources at large 
off-axis angles due to the increasing PSF size, which limits sensitivity.
\label{emap}}

\end{figure}

The current paper reaches to the low X-ray-to-optical flux 
ratios of Paper~VIII, but for the first time we are able to 
concentrate on the individually X-ray detected
objects.  This paper focuses on individually X-ray detected galaxies 
(of all morphological types) with 
\hbox{$\log{({{f_{\rm X}}\over{f_{\rm R}}})} < -2.3$} above a 0.5--2.0~keV 
flux limit of $\approx 2.3\times10^{-17}$~\flux (on-axis).  
One of the driving 
goals of this paper is to isolate a well-defined sample of 
 ``normal" galaxies within the X-ray source population for 
future comparisons (e.g., with \sirtf).  We have used an 
$\approx 12^{\prime} \times 15^{\prime}$ area within the 
2~Ms \hbox{CDF-N} survey \citep[][hereafter Paper~XIII]{davocatalog}, 
chosen because of its high effective exposure ($> 1500$~ks throughout 
most of the area; see Figure~\ref{emap}) and its overlap with
the Great Observatories Origins Deep Survey (GOODS) area where 
\hst\ Advanced Camera for Surveys (ACS) observations 
have recently been obtained  and \sirtf\ observations will be obtained later
this year.

In \S \ref{OBXF_data} we present the X-ray, optical, 
and near-infrared data used in this paper.  In \S \ref{sample_definition}
 we define the OBXF sample.  In \S \ref{opticalproperties_OBXF} we 
present optical spectra and describe the 
general optical properties of the OBXF sources, 
including their colors and morphologies.  
In \S\ref{xrayproperties_OBXF} we cover the X-ray 
properties of the OBXF population, including their luminosity 
distribution, X-ray spectral constraints, and X-ray number 
counts.  In \S\ref{discussion_OBXF} we discuss 
whether the OBXF galaxies are in fact ``normal" and 
investigate the nature of off-nuclear X-ray sources within the OBXF population.

The Galactic column density along this line of sight
is $(1.6\pm 0.4)\times 10^{20}$~cm$^{-2}$ \citep{Stark92}. 
An $H_0=70$~km~s$^{-1}$ Mpc$^{-1}$ and $q_0=0.1$ cosmology is adopted 
throughout this paper.  Coordinates are equinox J2000.


\section{Data and Source Detection}
\label{OBXF_data}

\subsection{Chandra ACIS Data and Source Detection}
\label{ACISobs_OBXF}

The X-ray data analysis performed here is essentially the same as performed
in Paper XIII.  In  this section we briefly describe the techniques.  The
data (described in Paper~XIII) were obtained during
20 separate observations by the \chandra\ Advanced CCD Imaging Spectrometer 
\citep[ACIS;][]{Weisskopf02,GarmireSPIE}  between 1999 Nov and 2002 Feb.  

The ACIS-I observations of the CDF-N cover a solid angle of 
462.3~arcmin$^{2}$.
However, the high-sensitivity region used in this study
only covers 178.9~arcmin$^{2}$ of the combined observation;
this region is hereafter referred to as the High-Exposure 
Area (HEA; see Figure~\ref{emap}). The $3\sigma$ flux limit in the central
part of this area is $2.3\times10^{-17}$~\flux (0.5--2.0~keV, see Paper~XIII), 
decreasing to $\approx3.0\times10^{-16}$~\flux at the outer edge of the HEA. 
Within this area, we can thus
detect galaxies with luminosities as faint as rest-frame $L_{\rm X}$(0.5--2.0~keV) 
$\simgt 8 \times 10^{38}$~erg~s$^{-1}$, $3 \times 10^{40}$~erg~s$^{-1}$, and 
$1.5 \times 10^{41}$~\lumin ~at $z=0.1$, 0.5, and 1.0, respectively.

\nocite{BrandtCatalog}

The average background over the HEA, excluding source counts, is
0.09~count~pixel$^{-1}$ in the 0.5--2.0~keV band with the standard ASCA
grades (Paper~XIII). All data were corrected for the radiation damage 
the CCDs have suffered and all X-ray spectral analysis was performed using the response 
matrices appropriate for such corrected data \citep{Townsley00,Townsley02}.
The fluxes have also been corrected for the molecular
contamination of the ACIS optical blocking filters.\footnote{See
http://cxc.harvard.edu/cal/Acis/Cal\_prods/qeDeg/ for more information on
the ACIS quantum efficiency degration.}

All of the X-ray sources were taken from Paper~XIII.
The X-ray bands considered are the 0.5--8.0~keV (full), 
0.5--2.0~keV (soft), and  2--8~keV (hard) bands.
The sources of Paper~XIII were detected above a  
{\sc wavdetect} \citep{Freeman02}
significance threshold of $1\times10^{-7}$ (see \S\ref{OBXF_significance}).
For sources that were not detected in a given band, upper limits are 
quoted at the $\approx99$\% confidence level.  
We have not corrected the X-ray fluxes 
for Galactic absorption; this effect is small due to
the low Galactic column density (see \S\ref{intro})
 along the line of sight. We have deviated slightly from the methods of Paper~XIII in
calculating X-ray fluxes as the average OBXF X-ray
spectrum implies a $\Gamma\approx2$ power-law spectral shape
(see \S\ref{stacked_spectrum_OBXF}) rather than the $\Gamma\approx1.4$
that is representative of the X-ray background \citep[e.g.,][]{miy98}.
In the cases where there are insufficient counts to constrain
an OBXF source's X-ray spectral shape,
we have thus assumed $\Gamma=2.0$
instead of $\Gamma=1.4$ (adopted in Paper~XIII).
The 90\% positional accuracy for the 
X-ray sources is $\approx0\farcs3$--1\farcs0.

\subsection{Optical and Near-Infrared Photometric Data}
\label{optphotomOBXF}

The optical and near-infrared photometric data 
used in this paper were drawn from several
databases.  For sources within the  
9\farcm0$\times$9\farcm0 Hawaii Flanking Fields region, 
we use the deep  $V$, $I$, and
$HK^\prime$ catalog of \cite{BargerHFF}. 
We have converted $HK^\prime$ to $K$ using the
$K = HK^\prime - 0.3$ relation of \cite{BargerHFF}.  

For the 17 X-ray sources that are outside the coverage area
of \cite{BargerHFF}, we have made use of deep optical $V$ and $I$-band images taken
with the Canada-France-Hawaii Telescope UH8K camera  
(G. Wilson et al., in preparation).
We used the {\sc sextractor} algorithm of \cite{Bertin96} with the
``Best" magnitude criteria, a $2\sigma$ detection threshold, and a
25-pixel Gaussian wavelet to measure magnitudes in the UH8K images.
The resulting sources were matched
to the catalog of \cite{BargerHFF} as a consistency check and were
found to agree within $1\sigma$ deviations of $\pm 0.1$ mag.
The 3$\sigma$ detection limits for these images are $V\approx26.3$ and
$I\approx25.1$.

We have adopted the $R$ band for sample definition and comparison
with other deep X-ray studies \citep[e.g.,][]{Schmidt98,Lehmann01}.
Where direct $R$-band measurements are not available, we convert 
between $V$,  $I$, and $R$ 
using the relation from Paper II:

\begin{equation}
 R = I - 0.2 + 0.5(V-I)
\end{equation}

For objects within the HDF-N and \hst\ Flanking Fields, we have also used
the \hst\ WFPC2 data in the f814w filter \citep[e.g.,][]{Williams96}.

\subsection{Source Significance for the OBXF Sample and the Parent Sample}
\label{OBXF_significance}

The effective significance of each OBXF X-ray detection is  
actually higher than the {\sc wavdetect} significance indicates
because we match X-ray sources to optically bright ($R \simlt 23$) 
counterparts, which are relatively rare on the sky 
\citep[for additional discussion, see \S 5.1 of ][and Paper XIII]{Richards98}.  
This allows us to adopt a matching radius slightly larger than the
0\farcs3--1\farcs0 positional accuracy of the $Chandra$ sources.
The reason for adopting a slightly larger matching radius is to search for possible off-nuclear 
X-ray sources \citep[these sources are fairly common locally and one has already been found in the
the CDF-N survey][hereafter Paper~I]{Ann00}.

As an example, at $R=22$ (the optically faint end of 
the population under study here) there are $\approx$13,000 
optically-selected galaxies ${\rm deg}^{-2}$  \citep{Steidel93}.   
Given a conservative X-ray/optical matching radius of 2\farcs0, 
the probability that any individual X-ray/optical match is 
false is $\approx 0.012$.  The effective significance 
threshold for a source detected with a {\sc wavdetect} 
significance of $1\times10^{-7}$ is thus really $\approx 1.2\times10^{-9}$.  
We conservatively estimate $<0.1$ false X-ray detections in total 
over the $\approx 2.6 \times 10^{6}$ pixels in the HEA.
For a more typical optical magnitude 
of $R=19.1$ (the median $R$ magnitude for the OBXF sample), 
the corresponding effective significance threshold 
is $< 7\times10^{-10}.$  

We have additionally evaluated the probability of false matches following
a method similar to that of Paper~XIII.  We shifted the positions of the 
X-ray sources by 5\farcs0 in both right ascension and declination (in four different
directions) and evaluated the number of X-ray/optical matches.  We found
that with a matching radius of 1\farcs0, we expect only one false match at $R<22$.
With a matching radius of 2\farcs0, this rises to four possible false matches,
and at 4\farcs0, we find that there should be $\approx14$ false matches.   We thus
adopt 2\farcs0 as our matching radius, but visually inspect all sources with 
offsets larger than 1\farcs0 to verify a likely association with a host galaxy.  We
have excluded three sources with 1--2\arcsec ~offsets because
the X-ray source's position falls outside the apparent optical
extent of the galaxy.

All X-ray
sources having an optical counterpart within 2\farcs0 are
considered as possible matches, but we visually inspect all objects with
$>$1\farcs0 offsets to verify a likely association with a host galaxy.  We
have excluded three sources with 1--2\arcsec ~offsets because
the X-ray source's position falls outside the apparent optical
extent of the galaxy.

There are a total of 293 soft-band sources detected above $2.3\times10^{-17}$~\flux (0.5--2~keV) within the
HEA (Paper~XIII).  In constructing a parent sample for comparison, we must also match to sources
that are more optically faint than the OBXF sample.  We adopt the same 2\farcs0 matching
radius for all sources with $R\leq 22$.  Following Paper~XIII, for sources with $R\leq23$ 
we adopt a 1\farcs5 matching radius.  For sources with $R>23$, we use a 1\farcs0 matching radius
or the positional error (whichever is larger).  Finally, for sources with $R>26.5$, we 
give an upper limit ($R>26.5$) as sources at these optically faint
magnitudes are not relevant to the current study.  For information on the optical properties of
these optically faint sources, consult A. Barger et al., in preparation.

Note that we do not use the supplementary catalog of
Paper~XIII which includes an optically bright, lower X-ray significance 
sample as the corresponding lower significance parent population 
contains too many false counterparts for reliable comparison.

\nocite{davofaint}

\section{Sample Definition}
\label{sample_definition}

\subsection{General Definition}
\label{general_definition}

Using the photometry in \S\ref{optphotomOBXF}, we define OBXF sources
to be those with $\log{({{f_{\rm X}}\over{f_{\rm R}}})} < -2.3$, 
where $f_{\rm X}$ is the X-ray flux in the soft band and 
$\log{({{f_{\rm X}}\over{f_{\rm R}}})}$ is calculated as follows:

\begin{equation}
\log{\left({{f_{\rm X}}\over{f_{\rm R}}}\right)} = \log{f_{\rm X}} + 5.50  + {R\over{2.5}}
\end{equation}

\noindent
The soft band was chosen for this definition to maximize sensitivity and because
the X-ray emission from quiescent galaxies is typically soft.
For instance, spiral galaxies, generally dominated by the emission from X-ray
binaries, have spectra reasonably well-approximated by
a $\Gamma=2.0$ power law \citep[e.g.,][]{Fabbiano95}.
The emission temperatures of the hot interstellar media of ellipticals are also
generally low, typically in the range $kT=0.5$--2~keV
\citep[e.g.,][]{Fabbiano95,IrwinPhD}.

However, to minimize contamination from hard X-ray emitting AGN 
we have added the additional requirement that the sources either have 
$\log{\left({{f_{\rm X}}\over{f_{\rm R}}}\right)} < -2$ in the 
full band or a hardness ratio constraint placing it softer than 
$\Gamma=1.0$.  We note that 
$\log{\left({{f_{\rm X}}\over{f_{\rm R}}}\right)} = -2$ 
using the full band corresponds to 
$\log{({{f_{\rm X}}\over{f_{\rm R}}})}=-2.3$ using the soft band 
if one assumes a $\Gamma=2.0$ power law.  With such a small 
number of photons, one cannot completely exclude the possibility 
of an obscured AGN but the current approach is fairly conservative.  
There are 15 sources with 
$\log{\left({{f_{\rm X}}\over{f_{\rm R}}}\right)} < -2.3$ in the 
soft band which fail this X-ray hardness filter.  

All of the sources are detected in the soft band.  We thus 
focus on the soft-band
properties of the OBXF sources as this is the band in which they are both
selected and individually detected.

\label{big_offsets_OBXF}

We have performed additional visual inspection of objects with optical
counterparts separated from the X-ray source by 
$2^{\prime \prime}$--$4^{\prime \prime}$, taking care to admit only
fairly optically bright galaxies as we might expect a large number of false matches
(see \S~\ref{OBXF_significance}).

\begin{figure}
\figurenum{2}
\centerline{\includegraphics[scale=0.90,angle=0]{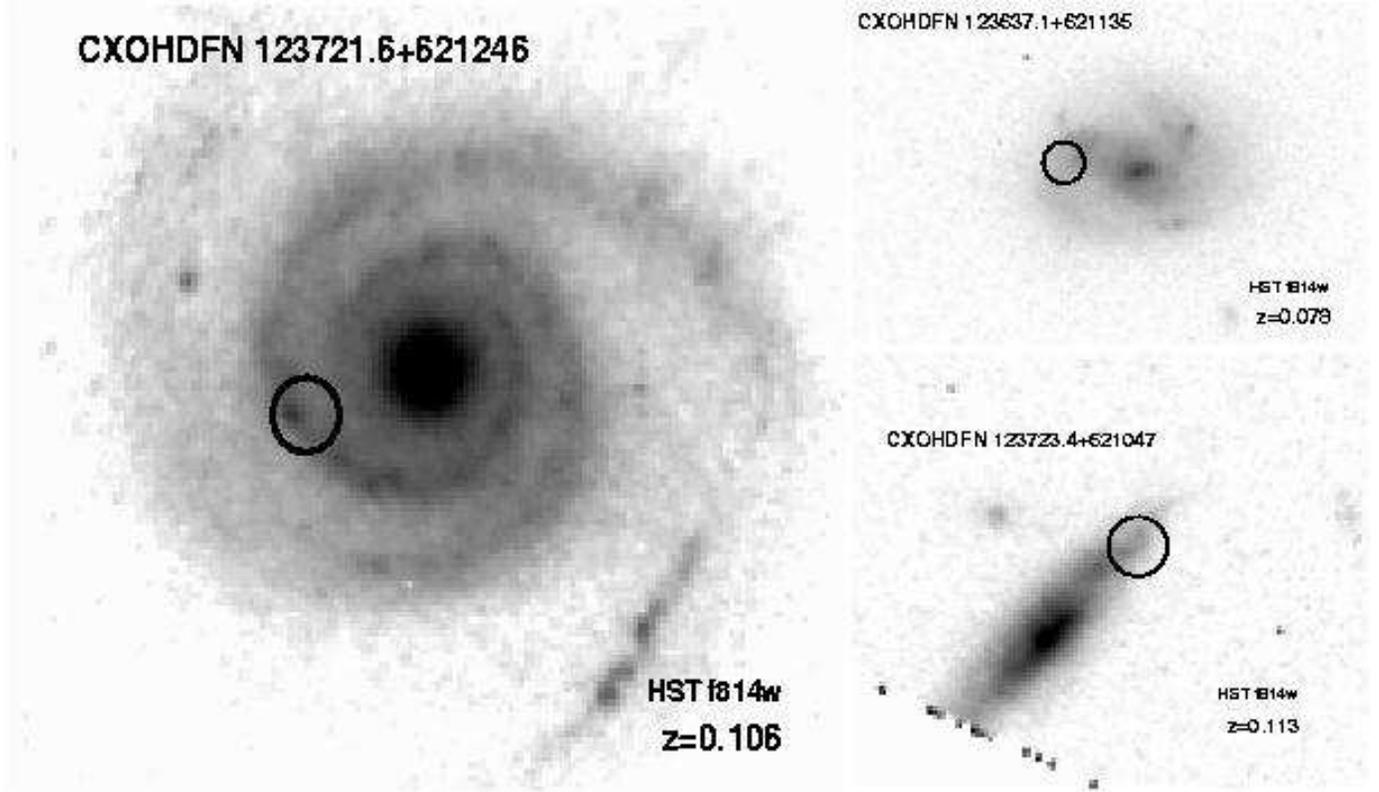}}

\caption[Optical cutout images, OBXF]{
HST f814w image  of the three OBXF sources with X-ray/optical offset $>2$\arcsec.
The location of the X-ray emission is marked by a circle with radius corresponding to the
X-ray positional error.  In CXOHDFN~J123721.6+621246, the X-ray source appears to be located
along a spiral arm and is coincident with a region of slightly enhanced optical emission.  In
the other two cases, association with either a spiral arm or the galaxy disk appears plausible.
\label{123721pretty}}

\end{figure}

We have been fairly conservative, finding that only three additional sources  
satisfy our visual inspection.
These three X-ray sources are all located within spiral/disk galaxy hosts,
have optical magnitudes $R \leq 19.6$ and X-ray/optical offsets $\leq$3\farcs0.
(see Figure~\ref{123721pretty}).   Had we arbitrarily matched sources having
$R \leq 19.6$ to X-ray sources with a $\leq$3\farcs0 matching radius, we would
have expected less than one false match (0.75 false matches).    

As an example, we discuss the X-ray source 
 CXOHDFN~J123721.6+621246 in more detail.  This source
is located in a grand-design spiral (see Figure~\ref{123721pretty}), along a 
spiral arm $\approx$2\farcs3 from the
galaxy's nucleus.  There is a slight optical enhancement at the
location of the X-ray source that is only apparent in the
\hst\ Flanking Field image; we estimate this knot to have
$I_{814}\approx25$--26.  The physical separation between the X-ray
source/optical knot and the galaxy's nucleus corresponds to
$\approx4.4$~kpc at the galaxy's redshift ($z=0.106$). Although the
formal chance that this is a random association is $\approx 2.8$\%, 
but the X-ray source is not randomly located
within the galaxy; a chance alignment with an optical knot in the
spiral arm is less likely.

\subsection{Summary of OBXF Sample Properties}
\label{OBXF_summary}

\begin{figure}
\figurenum{3}
\epsscale{1.10}

\centerline{\includegraphics[scale=0.50,angle=270]{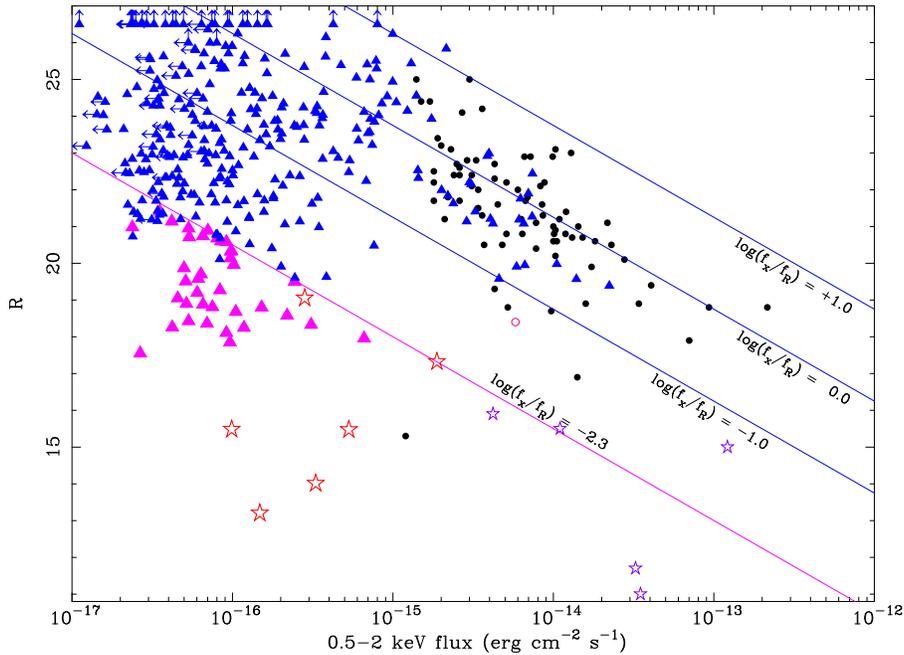}}

\caption[$R$ magnitude versus 0.5--2.0~keV flux]{
$R$ magnitude versus 0.5--2.0~keV flux. 
The OBXF sources are compared with the parent sample from
Paper~XIII (taken only from the HEA, see Figure~\ref{emap}).
Solid magenta triangles are the OBXF sources from this paper, 
and solid, slightly smaller blue triangles indicate the parent 2~Ms sample.
The seven red star symbols indicate the spectroscopically 
confirmed stars (see \S~\ref{zclass}).  The slanted lines 
indicate particular values of X-ray-to-optical flux ratio; the
lowest line marks \hbox{$\log{({{f_{\rm X}}\over{f_{\rm R}}})} = -2.3$}.
 Solid black circles mark the AGN from 
the \rosat\ Ultra-Deep Survey (UDS) of \cite{Lehmann01}. The open black 
circle is the one galaxy identified in the UDS, and the open black
star symbols are the UDS stars.  
We have excluded the groups of galaxies from \cite{Lehmann01} as we
only plot information for point-like X-ray sources.  
\label{fx_fR}}
\end{figure}

There are 43 OBXF sources with secure
X-ray/optical matches listed in Table~\ref{xraydata_OBXF}.
In Figure~\ref{fx_fR} we show a plot of $R$-band
magnitude  versus full-band X-ray flux for both the full 2~Ms X-ray
sample and the OBXF sources.

\begin{figure}
\figurenum{4}
\centerline{\includegraphics[scale=0.70,angle=0]{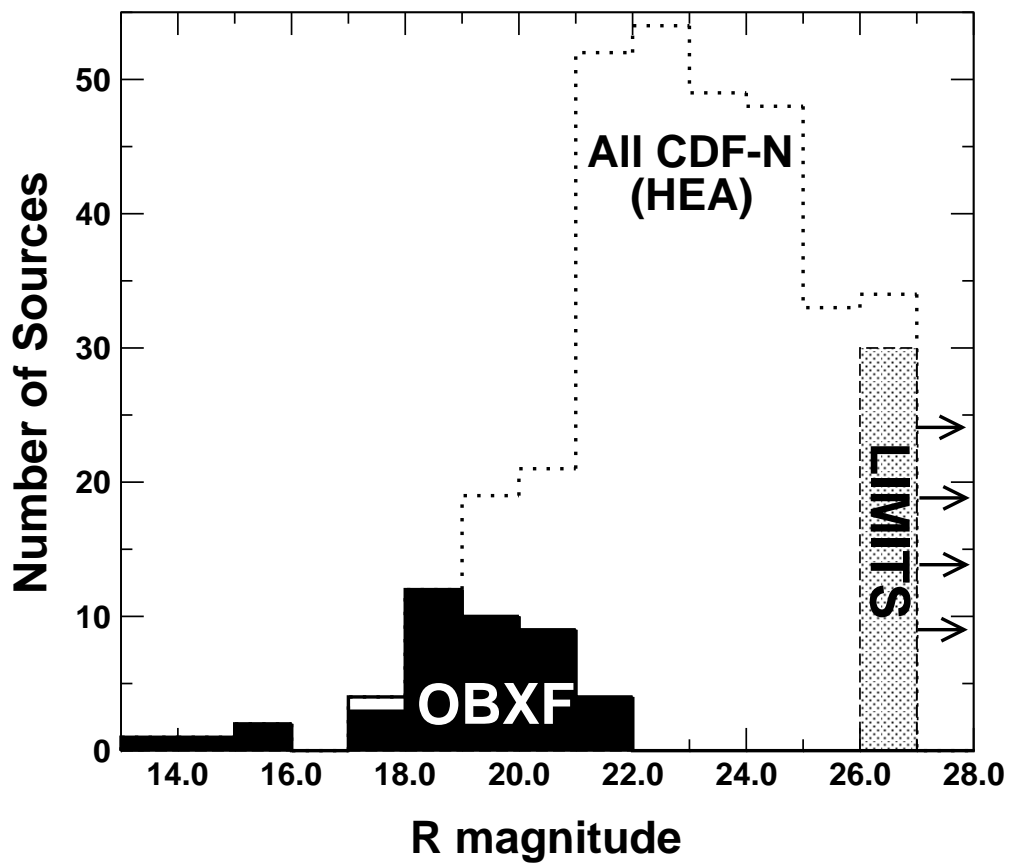}}

\caption[Optical Magnitudes of OBXF Sample]{
Distribution of $R$ magnitudes
for the parent 2~Ms X-ray sample (dotted histogram) and the OBXF sample (solid histogram).
The filled part of the OBXF histogram indicates sources for which we have
redshifts and/or identifications; only 1 source has neither.
The OBXF sources make up most of the brighter optical counterparts to
CDF-N sources.  
\label{opticalmagnitudes}}
\end{figure}

Figure~\ref{opticalmagnitudes} shows
the optical magnitude distribution for the OBXF sources and the
parent X-ray sample.  The median magnitude
of the 43 OBXF sources in the HEA is $R=19.1$.
The OBXF sources comprise the majority of the optically brightest X-ray
sources, but they clearly are still a minority of the overall X-ray
source population.   Figure~\ref{opticalmagnitudes} also shows
which OBXF sources have been identified through optical spectroscopy
or other means; there is only one unidentified source.

\begin{figure}
\epsscale{1.00}
\figurenum{5}
\plotone{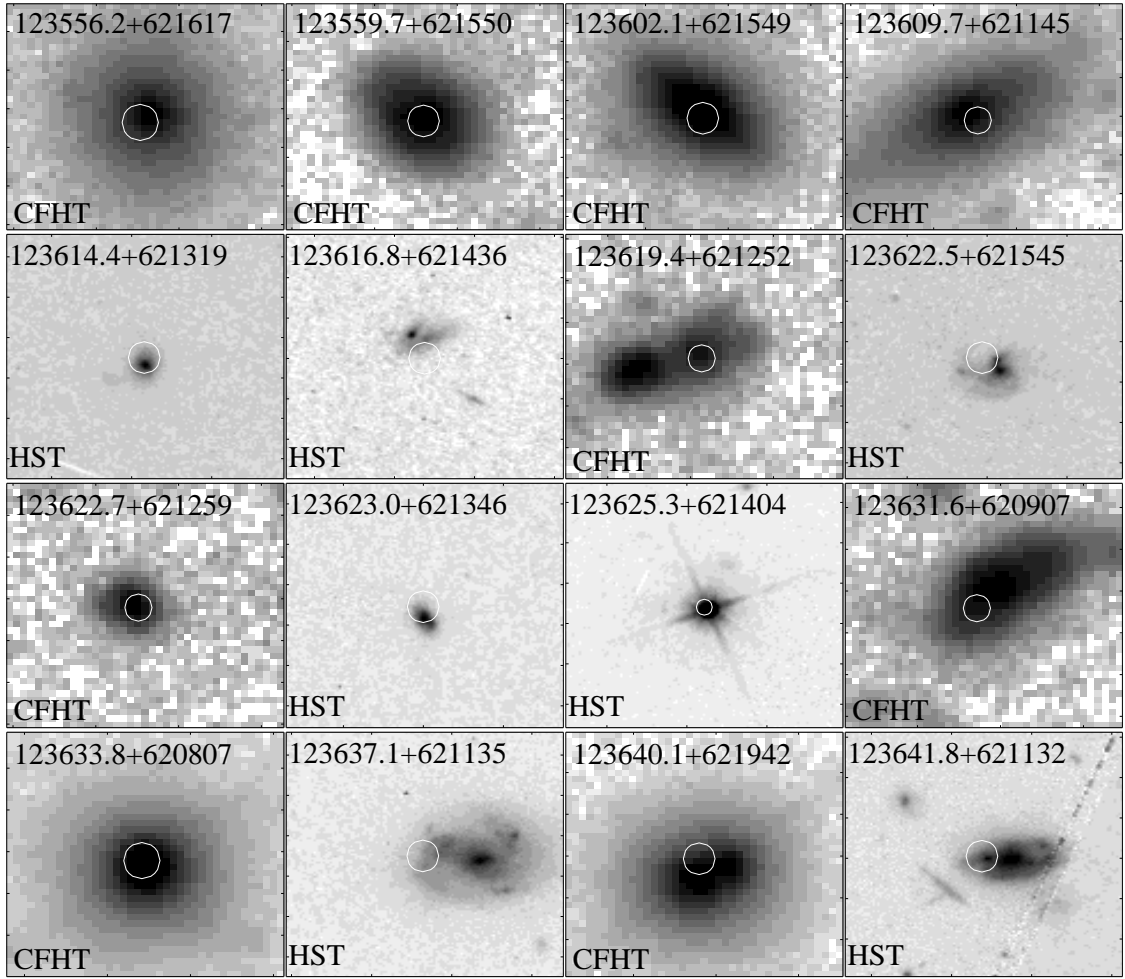}
\caption[Optical cutout images, OBXF]{Optical images of the OBXF sources.  
Each cutout image is $12^{\prime \prime} \times 12^{\prime \prime}$ in size.
The circles indicate the X-ray positional error as listed in Table~\ref{xraydata_OBXF};
the range of values is 0\farcs3--1\farcs0.  The upper labels give the IAU-registered 
CXOHDFN names, and the labels in the lower left-hand corner of each
panel indicate the telescope.  The filter for these images is the $I$-band.  
CFHT is the Canada-France-Hawaii Telescope and indicates data from G. Wilson 
et al., in preparation.  HST refers to the deep HDF-N and \hst\ Flanking-Fields 
data of \cite{Williams96}; for the HST images, the optical filter is f814w.
\label{cutout_images}}
\end{figure}

Optical images of the sources are shown in Figure~\ref{cutout_images}.
We display the ground-based $I$-band data described in \S\ref{optphotomOBXF}
except when the object lies in the HDF-N or \hst\ Flanking Fields.
In these cases, we display \hst\ imaging data 
from the f814w filter, which is most closely matched to
the $I$ band.  The f814w data are chosen as they cover both the \hbox{HDF-N}
and the Hubble Flanking Fields.

Figure~\ref{fraction_OBXF} shows the
fraction of X-ray sources that are OBXF as a function of X-ray flux.
This fraction increases at fainter X-ray fluxes, in
contrast to the optically faint ($I \simgt 24$) population which
is found to remain constant at $\approx35$\% of X-ray sources 
\citep[][hereafter Paper~VI]{davofaint}.
There are few OBXF sources in the sample at intermediate
X-ray fluxes ($\simgt 4\times10^{-15}$~\flux, 
0.5--2.0~keV), so the fraction there is less well-constrained. However, the 
fraction of OBXF sources at intermediate fluxes is roughly consistent 
with wider-field \chandra\ surveys \citep[e.g., $\approx5$\% of \chandra\ 
Multi-Wavelength Project (ChaMP) sources; ][]{Green03}.
Note that the 2~Ms data double the number of detected OBXF sources over
the 1~Ms data.

\nocite{Green02}

\section{Optical Properties of the OBXF Sample}
\label{opticalproperties_OBXF}

\subsection{Optical Spectroscopic Observations and Archival Redshifts \label{optspectra_OBXF}}

Redshifts and/or identifications have been obtained for 42 (98\%) of
the 43 OBXF sources using both the optical spectroscopy presented in this paper
and published spectroscopic redshifts.
There is only one object in our sample which is spectroscopically
unidentified, CXOHDFN~J123727.7+621034 ($R=18.0$).  
This source appears to reside in an optically resolved galaxy.
There are 25 optical spectra presented here;  
we collected 22, and three were obtained from archival resources.  
Of the 22 optical spectra we obtained, 15 are from the Hobby-Eberly
Telescope (HET), three are from Kitt Peak, and four are from Keck.
Here we briefly describe the optical spectra 
and archival redshift sources.

The Keck Low Resolution Imaging Spectrograph \citep[LRIS; ][]{Oke95}
was used to observe OBXF sources during 2001 March.
The integration times were typically one to two hours.
The blue side of LRIS was in operation at this time; some of these 
spectra extend down to $\approx3000$~\AA ~with an occasional gap in
spectroscopic coverage at 5600~\AA .  Where possible, we have normalized
the blue spectrum to match the flux of the red spectrum in the region
of overlap.  Three of the Keck LRIS observations were obtained 
in multi-slit mode and one (CXOHDFN~J123641.8+621132) was obtained in 
long-slit mode.  We obtained a long-slit spectrum of 
\hbox{CXOHDFN~J123641.8+621132}. The latter's X-ray emission is offset from
the host galaxy nucleus and is coincident with an ultraviolet-bright knot.   
We placed the slit on this knot to determine if it is a background AGN and
verified that the knot also lies at the $z=0.089$ redshift determined
for this galaxy by \cite{Cohen00}.

Spectra were also obtained using the Marcario Low Resolution
Spectrograph \citep[LRS;][]{HillHET98a,HillHET98b,Schneider00} of the
HET \citep{Ramsey98}. The HET operates in
a queue-scheduled mode, so the observations occurred on 
multiple evenings between 2000 February 7 and 2001 June 2.
  A 2\farcs0 slit and a 300~line~mm$^{-1}$ grism/GG385
blocking filter produced spectra from 4400--9000~\AA\ at
17~\AA\ resolution; data redward of  7700~\AA\ are suspect because of
possible second-order contamination.  The exposure time per source
ranged from 5--60~minutes.
 The seeing was typically 2\farcs5 (FWHM).
Wavelength calibration was performed using HgCdZn and Ne lamps, and
relative flux calibration was performed using spectrophotometric
standards.

The other spectra were
obtained during multi-slit observations with the R.C. Spectrograph 
on the Mayall 4~m telescope at Kitt Peak the  nights of 2002 May 15 and May 16.
We used the BL400 158~line~mm$^{-1}$ grating  and the OG530 blocking filter
to produce spectra from 5300--10000~\AA\ at 14.5~\AA\ resolution.  
Wavelength calibration was performed using HeNeAr lamps, and relative 
flux calibration was performed using spectrophotometric standards.  

Two Keck LRIS spectra are included from the publicly available database of 
\cite{BargerHFF}\footnote{http://www.ifa.hawaii.edu/$\sim$cowie/hdflank/hdflank.html}. Two spectra from the publicly available Deep Extragalactic 
Evolutionary Probe \citep[DEEP, e.g.,][]{KooDEEP} database of 
Keck spectra are also included.\footnote{Both the LRIS and ESI 
instruments were used.  See http://saci.ucolick.org/verdi/doc/spectra.html} 
The band pass for these Keck LRIS observations 
is $\approx 5000$--10000~\AA.  

For the remaining objects, redshifts and spectral identifications are included from
\cite{BargerCatalog2002}, A. Barger et~al., in preparation; and 
\cite{Cohen00}.  

\subsection{Redshifts and Classifications from Optical Spectra \label{zclass}}

The optical matches and identifications for the OBXF sample are given
in Table~\ref{multiwav_OBXF}.  
In Figure~\ref{spectra1} we show the 26 optical spectra of OBXF
sources (see \S\ref{optspectra_OBXF}).  A total of five
previously unpublished redshifts are presented; all of these 
lie outside of the HDF-N and \hst\ Flanking Fields.
All the spectra are plotted in terms of  rest-frame wavelength.  

Only broad spectral classifications of the sources 
are made due to the low signal-to-noise of some of the spectra.  
For the 16 objects for which we did not have a spectrum, we relied on the
spectral classifications of \cite{Cohen00} or classified the optical
spectra of either \cite{BargerCatalog2002} or A. Barger et al., in preparation.
Our classifications, similar to those of \cite{Cohen00}, are  
``absorption-line dominated" (${\cal A}$), 
``composite" (${\cal C}$), ``emission-line dominated" (${\cal E}$),\footnote{
Objects of \cite{Cohen00} type ``intermediate",
``intermediate/emission-dominated", and
``emission-dominated/intermediate" were classified as ${\cal E}$ here.}
and ``stellar" (${\cal S}$).   Objects of composite type 
exhibit a strong [\OII]~$\lambda 3727$ emission line as well as the Balmer 
series in absorption. In a few cases we have obtained spectra 
of objects classified as ``intermediate" by \cite{Cohen00} 
and determined that they were of type ${\cal C}$.  
Only objects of \cite{Cohen00} 
class ${\cal A}$ were placed into our absorption-dominated class.  
We cast all stars \citep[${\cal M}$ and ${\cal S}$ of][]{Cohen00} 
into class ${\cal S}$.   

Through our optical spectroscopic identifications, we 
find that 9 sources are absorption-line dominated (21\% of identified
sources), three sources are composite (7\%), 24 sources are
emission-line dominated (57\%), and six sources are stars (14\%).  

The emission-line dominated sources are a diverse group, 
with some exhibiting emission only in [O~{\sc ii}]~$\lambda 3727$  
and others showing clear emission in both 
[O~{\sc iii}]~$\lambda 5007$ and H$\beta$.  
Among the emission-line dominated galaxies where we have optical
spectra, we find that 
a significant fraction have strong [O~{\sc iii}]~$\lambda 5007$ 
and H$\beta$ emission, with the rest exhibit 
[O~{\sc ii}]~$\lambda 3727$ and in some cases Ca H and K 
absorption.  In a few very low redshift cases, H$\alpha$ 
is also seen.

There are six objects identified as stars. 
Six of these are mid-to-late type stars (M4-5), 
and one is an F star (CXOHDFN~J123737.9+621631).

The redshift distribution of the identified extragalactic sample 
(38 sources) is shown in Figure~\ref{z_distribution}.
Of these 38 sources, $\approx 97$\% have $z < 1.0$, $\approx 84$\% have 
$z < 0.5$, and the median redshift is $z=0.299$.  Note that there
is a possible ``peak" in the number of galaxies in the 
$0.4 \simlt z \simlt 0.6$~ redshift interval \citep[see the similar findings
in ][]{BargerCatalog2002}.  Four of the nine
galaxies in this interval are in a very narrow redshift range,
$0.409 \simlt z \simlt 0.412$, which corresponds to the redshift range 
for one of the previously determined overdensities of optically-selected 
galaxies \citep{Cohen00}.   

\subsection{Optical Colors \label{opticalcolors_OBXF}}

$I-K$ colors of the 24 OBXF sources that lie within the deep 
multi-band imaging region of the Hawaii Flanking Fields 
\citep{BargerHFF} are shown in Figure~\ref{plot_colors_OBXF}. 
Figure~\ref{plot_colors_OBXF} also shows the color distribution of the
field galaxy population \citep{BargerHFF}. 

There is a trend in optical/near-infrared colors for 
X-ray sources found in deep surveys 
\citep[excluding broad-line AGN; e.g., Paper II;][Paper 
VI]{Hasinger98,Lehmann01,Giacconi01}.  The optical counterparts 
to X-ray sources have among the reddest optical/near-infrared 
colors within the optically 
selected source population at a given optical magnitude. 
The OBXF population is distributed more uniformly with respect to $I-K$ colors and $I$-band
magnitude (see Figure~\ref{plot_colors_OBXF}).  These ``bluer" 
red/infrared colors of the OBXF population are more consistent 
with the field galaxy population.  

We compare the OBXF optical colors to the work of \cite{Koeke02}, who
have performed an \hst\ imaging study of the \chandra\ Deep Field-South
\citep[hereafter CDF-S; e.g.,][]{Giacconi02}.  This study included
three \hst\ WFPC2 fields  and contained analysis of 40 X-ray sources, regardless
of optical brightness/faintness.  
\cite{Koeke02} made the broad statement that the optically bright population
has redder colors than the field galaxy population; this conclusion
was based on ${\rm f606w}-{\rm f814w}$ (approximately $V-I$) colors.  However, they note
the presence of a bright class having ``intermediate" colors which are generally
associated with optically resolved galaxies.   These 
intermediate sources have colors consistent with the OBXF sample.

\subsection{Host Galaxy Morphology}
 \label{OBXF_morph}

To address the issue of host galaxy morphology, we consider the \hst\
imaging data for the 18 OBXF galaxies within the HDF-N and \hst\  
Flanking Fields (see Table~\ref{morph_table_OBXF}).  

The morphological classification follows the quantitative 
measuring technique of \cite*{cc00} and \cite*{bershady00}.
This system is  not based on the Hubble sequence, 
although we do discuss the morphological parameters in the context of the Hubble
sequence.  The strength of this quantitative morphological 
analysis is that it is robust
for both distant and nearby galaxies.  The morphological parameters
are measured on images taken in a given optical filter; for example
we write the asymmetry index as ``$A(I)$" because
the morphology is measured in the f814w band, which is closely matched
to the $I$-band.   

The asymmetry index, $A$, reveals the distribution
of light and gives some indication of the possible
distribution of the matter in the galaxy.
This index is calculated by rotating the galaxy image ($I_{0}$) through
different values of the angle, $\phi$, and measuring the
residuals after subtracting the original image.  The
parameter is normalized such that when $A=0$, the
galaxy is symmetric to all rotations ($I_{0}-I_{\phi}=0$),
and when $A=2$ the galaxy is completely asymmetric so that
at a given $\phi$, $|I_{0}-I_{\phi}|=I_{0}$. The asymmetry
index reveals the dynamical processes in galaxies, measuring
the degree of recent interactions/mergers.

The concentration index ($C$) effectively measures the
radial distribution of light and may give some indication
of how a galaxy was assembled as well as its scale. 
If the number is high (such as in an elliptical, 
$C \simgt 3.5$) then it is thought that the galaxy was produced 
through  dissipationless mergers.   Low concentrations are 
usually found in disk-dominated systems that have high angular 
momenta and thus do not have as concentrated light.

Table~\ref{morph_table_OBXF} lists $A(I)$ and $C(I)$ for all
of the OBXF galaxies in the \hbox{HDF-N} and HST Flanking Fields.  We 
also include some notes on individual sources, including
visual assessments of the host galaxy morphology.
Note that all of the galaxies identified as early-type/ellipticals
by eye are in a transition stage, having slightly higher
than average asymmetry.  Comparison with the more traditional
Hubble types (broadly divided into late-type, intermediate,
and early-type galaxies) is made using the 
the calibration of \cite{bershady00} for this algorithm on
nearby well-resolved galaxies.  

We divide the OBXF galaxies into two redshift intervals ($z<0.5$ and $z>0.5$)
in order to compare the objects 
with \hbox{HDF-N} field galaxies measured using the
same \cite{cc00} algorithm \citep[where \hbox{HDF-N} spectroscopic redshifts were not
available, we used the photometric redshifts of][]{Budavari00}.  
We compared the OBXF galaxies with reasonably optically luminous galaxies;
we only calculate the morphological parameters for field galaxies with
$M_{\rm B} < -18$.  The values of the measured parameters are plotted in
Figure~\ref{MORPH_OBXF}.  There are some qualitative differences in
the OBXF host galaxy morphologies as compared to the field population;
at lower redshifts the OBXF galaxies are 
of later type than at higher redshifts.  Also, there are no OBXF 
galaxies of classical early-type.

We find that the concentration indices are virtually
identical for the OBXF and field samples ($<C>=3.1$--3.5).
There is an indication that the  OBXF sources at $z>0.5$ may be
slightly more asymmetric than the field galaxies: 
$<A>=0.19$ for the field galaxies and $<A> = 0.45$ for the OBXF galaxies 
but the scatter in the OBXF values ($1\sigma$ scatter in
$<A>$ is $\approx0.24$ at $z>0.5$) 
 is fairly large so they are statistically consistent.  

Due to the likely connection between X-ray emission and
vigorous star-formation resulting from mergers,
it is expected that one might find more instances of mergers/interactions 
among the OBXF sources than the
field galaxy population.  Using $A = 0.35$  as a 
merger limit \citep{cc00b}
 it appears that three ($\approx17$\%) of the OBXF sources 
are undergoing or recently underwent a merger.  
This is slightly higher than the value of 10\% found for field galaxies over the same
redshift interval. To compare these fractions, we perform a Fisher exact probability
test for two independent samples \citep[one is the sample
of 208 optically selected galaxies, the other is the sample of 18 X-ray selected
OBXF galaxies; e.g., ][]{Siegel88}.  We find that
the probability that X-ray selection and optical selection 
of galaxy mergers is different is $\approx85$\%.  The statistical
level of the difference is thus marginal, although it does 
leave open the possibility that deep X-ray surveys
preferentially select galaxies undergoing mergers as compared
to the general field galaxy population.  

Since $Chandra$ has excellent spatial resolution, one might also consider the
X-ray morphology of the galaxies.  This is difficult in most of the cases due
to the low number of X-ray counts;  only one of the OBXF sources has any hint of X-ray extent.  This
one source is near the aimpoint of the observations.  We thus suspect that this
apparent extent could be due to the very complicated CDF-N coadded PSF.

We can compare our findings with those of the CDF-S. In Table~4 of \cite{Koeke02}
there are ten elliptical galaxies, but nine of these ten galaxies have 
higher X-ray-to-optical flux ratios than are being considered here.  The tenth object
would have been marginally within our selection criterion.  The lack of elliptical
galaxies in the current sample thus appears consistent with findings in the CDF-S.

\section{X-ray Properties of the OBXF Population}
\label{xrayproperties_OBXF}

Figure~\ref{Lx_histogram} compares the
X-ray luminosity distribution of the OBXF sample with a 
sample of normal late-type galaxies detected in the local Universe by
\einstein\ (Shapley et al. 2001).  Also plotted in Figure~\ref{Lx_histogram}
are the X-ray luminosities of the Milky Way \citep[$\approx4\times10^{39}$~\lumin, 
0.5--2~keV,][]{Warwick02}, 
the local starburst M82 \citep{Griffiths00}, and the infrared-luminous 
starburst NGC~3256 \citep[the most X-ray-luminous nearby starburst known; ][]{Moran99,Lira01}. 
There are clearly a large number of
sources in the OBXF sample with X-ray luminosities  
higher than those of normal late-type galaxies.
Figure~\ref{Lx_distribution}a shows that the luminosities,
not unexpectedly, are a strong function of redshift.  The 
expected luminosity sensitivity near the \chandra\ aimpoint
is also plotted in Figure~\ref{Lx_distribution}a, and we see
that even with 2~Ms of exposure, \chandra\ can only 
detect galaxies with the low X-ray luminosities comparable to
the Shapley et~al. sample ($L_{\rm X} \approx 10^{40}$~\lumin) 
at $z\simlt0.3$.  At higher redshifts, it appears that the
galaxies being probed are luminous starbursts
and moderately luminous AGN.  Figure~\ref{Lx_distribution}b shows that
the lower luminosity, nearby objects also have  
lower X-ray-to-optical flux ratios and that we are sampling a much
greater spread of low X-ray-to-optical flux ratios.  
We make a division between the ``high group" and the ``low group" 
to determine the X-ray properties of the sample.  The high group consists of
14 galaxies having $\log L_{X} \simgt 40.5$ and 
$\log{({{f_{\rm X}}\over{f_{\rm R}}})} \simgt -2.5$.  The low group
consists of 22 galaxies having $\log L_{X} \simlt 41.0$ 
and $\log{({{f_{\rm X}}\over{f_{\rm R}}})} \simlt -2.5$.  These two
groups are marked in Figure~\ref{Lx_distribution}b.

\subsection{Stacked X-ray Spectrum of Normal Galaxies 
\label{stacked_spectrum_OBXF} } 

The small number of X-ray counts per OBXF source
(the median number of full-band counts is
35.4) prohibits extensive X-ray spectral analysis on a source-by-source basis.
There are only five sources detected in the hard band and of these, only three
have band ratios which imply fairly hard X-ray spectra.
These sources are described in detail in the appendix;
they appear to have just narrowly passed our selection
criteria as OBXF sources. They  are more typical of starbursts and/or lower-luminosity
AGN.  

We have stacked the X-ray counts from each of the OBXF galaxies
to make a combined spectrum to constrain further
their X-ray spectral nature and their contribution to the
X-ray background.  The average response matrix
and ancillary response file (arf) are calculated as
described in \S\ref{ACISobs_OBXF}. 
We include the 36 galaxies for which we have
redshifts and/or spectroscopic identifications.
We have excluded data above 
5~keV because of insufficient counts.  The extraction 
regions are circles of radius 4 $Chandra$ pixels.

Figure~\ref{xray_spectrum_OBXF} shows the composite
0.5--5.0~keV spectrum for these 36 OBXF sources. 
The  0.5--5.0~keV spectrum contains a 
total of 1526 counts; 447 (29\%) of
these are expected to be from the background.  
The power-law photon index of this spectrum, assuming the 
Galactic column density, is $\Gamma=1.98^{+0.16}_{-0.15}$ 
(95\% confidence errors, reduced $\chi_{\nu}^{2} =1.05$,${\nu}=61$).   
The integrated 0.5--2.0~keV flux of the OBXF galaxies is 
$2.2\times10^{-15}$~\flux.  This is $\approx1$\% of the 0.5--2~keV 
cosmic X-ray background (XRB) assuming a spectral form for the XRB
of $10 E^{-1.4}$~keV~keV$^{-1}$~cm$^{-2}$~s$^{-1}$~sr$^{-1}$
(see Vecchi et~al. 1999 for a discussion
of recent efforts to measure the intensity of the XRB).
\nocite{Vecchi99}

We have split the OBXF sources into optical spectral classes and
considered their X-ray spectral properties.  Both the ${\cal E}$ and
${\cal A}$ galaxies are reasonably well fit by a $\Gamma=1.8$ power-law, which
is reasonably consistent with what has been found for local galaxies
\citep[e.g.,][]{Fabbiano95}. 
We have also considered the high group and low group as defined
earlier in this section.  The stacked X-ray spectrum of the low group 
may be described by a power-law with $\Gamma=1.90^{+0.27}_{-0.26}$ (95\% confidence errors, 
reduced $\chi_{\nu}^{2} =0.68$,${\nu}=47$).  For the high group, the
X-ray spectrum is not well-fit by any simple model; 
the best fit obtained was for $\Gamma=2.14$ ($\chi_{\nu}^{2} =1.26$,${\nu}=33$).
This agrees reasonably well with the $\Gamma\approx 2$ spectrum
obtained previously from stacking a smaller sample of X-ray detected
infrared-luminous galaxies (11 galaxies; \S 3.4 of Paper~XI).

The OBXF galaxies do not divide themselves simply
by X-ray spectral properties, but this is not so unexpected as they vary quite a bit
even within these subgroups in terms of properties such as morphology and
luminosity.  The overall soft nature does suggest there are very few obscured
AGN within the population.

\subsection{X-ray Number Counts}

We have compared the number counts of the extragalactic
OBXF population with those of the full CDF-N X-ray source population
(see Figure~\ref{logNlogS}).  The low X-ray fluxes of the OBXF
population lead to a fairly narrow range of X-ray flux over
which number counts may be reliably well-measured.  The sensitivity maps
derived for Paper~XIII were used to determine the area over
which we are  sensitive to sources of a given flux.    We have excluded
all sources having soft-band fluxes less than
$4.1\times10^{-17}$~erg~cm$^{-2}$~s$^{-1}$
due to incompleteness below these detection levels.
There are insufficient numbers of sources above 
$1.7\times10^{-16}$~erg~cm$^{-2}$~s$^{-1}$ to 
constrain the number counts, so this is adopted as an upper bound.
There are 29 OBXF sources between these flux levels.

A maximum likelihood fit, assuming a power law \citep[e.g.,][]{Murdoch1973},
was made to the soft-band differential number
counts from $4.1\times10^{-17} $~erg~cm$^{-2}$~s$^{-1}$ to
$1.7\times10^{-16} $~erg~cm$^{-2}$~s$^{-1}$. The fit 
yielded a slope of $-1.46^{+0.28}_{-0.30}$ (90\% confidence) 
for the corresponding cumulative number counts.  The number
counts are described by the following equation ($N(>S)$ is the number
of sources per square degree).

\begin{equation}
N(>S)=310\left({S\over 1\times 10^{-16}}\right)^{-1.74^{+0.28}_{-0.30}}
\end{equation}

By comparison, the slope for the general soft-band detected
X-ray source population is quite flat over the same flux range
at $-0.67\pm0.14$  (Paper~V).

Indirect measures of galaxy number counts, which have been 
able to probe galaxies statistically beyond the formal 1~Ms detection limit, 
have included the stacking analysis work of Paper~VIII, which 
focused on quiescent spiral galaxies, and the fluctuation 
analysis work of, e.g., \cite{Miyaji02}.  In Figure~\ref{logNlogS} 
we show these results; an extrapolation of the OBXF normal
galaxy counts should intercept the Miyaji fluctuation analysis ``fish"
at a 0.5--2~keV flux of $\approx7\times10^{-18}$~\flux.     This is
a coarse estimate of the flux where the X-ray number counts will
be dominated by normal galaxies and is in reasonable agreement
with the estimates of \cite{Ptak01} based on the optical properties
of field galaxies (also shown in Figure~\ref{logNlogS}) and of
Paper~VIII.

\section{Discussion}
\label{discussion_OBXF}

\subsection{Truly ``Normal" Galaxies? \label{trulynormal}}

It is only by extending the CDF-N survey to 2~Ms  and covering
an area larger than the HDF-N and its immediate environs that
we have obtained a sizeable sample of fairly normal galaxies.
In this section we revisit the question of just how ``normal" the
galaxies are that are appearing in deep \chandra\ surveys, in particular
to establish that they are consistent with X-ray emission from normal
galaxies rather than highly obscured AGN.  It is thus important to
point out that the galaxies presented here are
completely different in nature from the ``optically dull", X-ray
luminous galaxies found in the HELLAS survey \citep[e.g.,][]{Fiore00}.  
\cite{Comastri02} makes the good suggestion to refer 
to these X-ray luminous galaxies as XBONGS (X-ray Bright, Optically Normal Galaxies) 
to make the distinction clear.  In this paper ``normal" refers to the
X-ray and optical emission from the galaxy, normal can be equated in
a {\it very} coarse sense with ``Milky Way-type" galaxies. 

We find that even with the fairly conservative X-ray-to-optical 
flux ratio cutoff imposed here, a significant fraction of 
sources have X-ray luminosities of $\approx10^{41}$~erg~s$^{-1}$ 
(see Figure~\ref{Lx_histogram}) rather
than the $10^{39}$--$10^{40}$~erg s$^{-1}$ that is more typical of normal
galaxies in the local Universe \citep[e.g., ][]{Shapley01}.
Figure~\ref{Lx_distribution}a shows that this is basically a
sensitivity effect as we are only able to detect normal
galaxies to $z\approx0.3$ even with 2~Ms of
\chandra\ exposure.   This is consistent with the findings of Paper~XI that  
starburst galaxies dominate over normal (less X-ray luminous) galaxies 
even with 1~Ms of exposure.

We find that even at the faint X-ray flux limits of a 2~Ms \chandra\
survey, the X-ray-to-optical flux ratio is
a useful discriminator of galaxies versus AGN  \citep[similar to
the findings of Paper~XI and earlier results at much brighter
X-ray fluxes, e.g., ][]{Macc88,Stocke91}.
Consider the relationship between X-ray luminosity and
${f_{\rm X}}\over{f_{\rm R}}$  in Figure~\ref{Lx_distribution}b. 
There are no extremely X-ray luminous OBXF sources.

Overall, $\approx12$\% of X-ray sources in 2~Ms surveys 
are ``normal" galaxies.  Table~\ref{sources_by_fxfR} 
gives a breakdown of the source types by full band X-ray-to-optical 
flux ratio.  Also included in Table~\ref{sources_by_fxfR} is
the breakdown for sources with 
slightly higher values of X-ray-to-optical flux ratio
[$-2 < \log{({{f_{\rm X}}\over{f_{\rm R}}})} <-1$; Paper~XI], we
have extrapolated these 1~Ms identifications to the 2~Ms sample. 
We find that while AGN certainly do dominate as far as numbers of
sources in deep (1--2~Ms) surveys, 10--30\% of the sources may be explained
through high-energy emission processes other than accretion onto
supermassive black holes.  The range in values is due to the 
unknown prevalence of lower-luminosity AGN in these sources.

One can ask the reciprocal question concerning what fraction of
 ``normal" galaxies are being detected in 2~Ms X-ray surveys.  
In the CDF-N, we expect to be able to detect galaxies 
having $L_{\rm X} = 10^{39.8}$ erg s$^{-1}$ to $z\approx0.3$ 
and to detect galaxies having $L_{\rm X} = 10^{39.2}$~erg s$^{-1}$ to
$z\approx0.15$.  We have considered galaxies within the 
8\farcm6 $\times$ 8\farcm7 Caltech Faint Field
Galaxy Redshift Survey Area (hereafter the Caltech Area)
because its redshift coverage is highly complete \citep{Cohen00} 
and because there are $B$-band luminosity calculations 
available for this region \citep{Cohen01}.  We detect  all six of the
galaxies with $z < 0.15$ and $M_{\rm B} < -19$ in the \cite{Cohen00}
sample as OBXF sources.\footnote{
$M_{\rm B} = -19$ corresponds to $0.3L_{\rm B}^{\ast}$; we have used the
equations of \cite{Cohen01} to calculate $M_{\rm B}$ and the Sloan Digital Sky
Survey value for $M_{\rm B}^{\ast}$ \citep{Blanton01}.}  
This number of low-redshift detections
is consistent with that found over the rest of the HEA.
We detect few of the galaxies in the Caltech Area
having $0.15 < z < 0.3$ and $M_{\rm B} \simlt -19$; constraints
from the rest of the HEA OBXF sample (i.e., outside the Caltech Area)
indicate that $\approx 14$\% of galaxies with $0.15 < z < 0.3$ are
detected.

We thus have the sensitivity to detect all optically luminous
galaxies (non-dwarfs) out to $z\approx0.15$
and some ($\approx14$\%) out to $z\approx0.3$.  This is roughly 
consistent with the early findings in Brandt et al. (2001; hereafter Paper~IV);
\nocite{BrandtIV}
 the current study has much better statistics on the fraction 
of optically normal galaxies detected.

\subsection{Off-nuclear Sources \label{ULX_OBXF}}

Five of the OBXF sources are candidate off-nuclear X-ray sources
and are listed in Table~\ref{ULX_OBXF_candidates}. All of these objects
have full-band X-ray luminosities $\simgt 10^{39}$~\lumin, indicating
they are members of the off-nuclear ultraluminous X-ray (ULX) population
that was discovered during the \einstein\ era \citep[e.g.,][]{Fabbiano89}.
Their X-ray luminosities typically exceed that expected for 
spherically symmetric Eddington-limited accretion onto ``stellar" mass
(5--20~\msun) black holes\footnote{The Eddington-limited luminosity for
a 15~\msun black hole, assuming spherically symmetric accretion, is 
$2.0\times10^{39}$~\lumin.  All but one of the sources presented here exceed
this luminosity.}. They are thus ``ultraluminous" as compared 
to other sub-galactic X-ray emitters but are
of course very low luminosity compared to luminous cosmic X-ray emitters (AGN).
At $z\simgt0.1$, the high angular resolution
of \chandra\ is vital in order
to establish that the sources are indeed off-nuclear.  Complementary
sub-arcsecond observations with \hst\ are also needed
to verify association with bright HII regions or
spiral structure within the host galaxy.  Thus, the 2~Ms CDF-N data present a unique
opportunity to study these objects up to look-back times of \hbox{$\approx 1$--2}
 gigayears.

ULX sources may still be consistent with stellar mass black holes, possibly
representing an unstable phase in normal high-mass X-ray
binary evolution \citep[e.g.,][]{King01} or possibly a population of
rapidly spinning Kerr black holes allowing higher X-ray luminosities
\citep[e.g.,][]{Makishima00}.  They also possibly represent a class of
intermediate mass black holes \citep[$\approx 500$--1000~\msun, e.g., ][]{Colbert2002}
or ultraluminous supernova remnants
\citep[e.g.,][]{Blair01}.  While \chandra\ and \xmm\ have recently been
able to study these sources in much greater numbers in nearby galaxies
\citep[e.g.,][]{Roberts02} and \rosat\ observations of samples of nearby
bright galaxies are producing large numbers of candidates\citep{Colbert2002},
there is still fairly little information about how
common these sources are throughout the Universe.

Two of the CDF-N OBXF ULX sources are consistent with 
bright HII regions displaced from the nuclei of the host
galaxies (grand-design spiral CXOHDFN~J123721.6+621246; 
mentioned in \S \ref{big_offsets_OBXF} and the previously 
published off-nuclear ULX CXOHDFN~J123641.8+621132,
located in the HDF-N itself; Paper~I).
These two sources are described as ``highly confident" candidates
in Table~\ref{ULX_OBXF_candidates} due to their likely association with
structures within their host galaxies.  We have identified three  
additional candidate off-nuclear X-ray sources which are also listed in
Table~\ref{ULX_OBXF_candidates}.

The two most confident ULX candidates show strong signs of
variability and several of the other candidates show evidence for
variability.  This variability, along with the high X-ray luminosities,
 indicates that these sources are black hole candidates.
 For more detailed information on variability in ULX sources
at these redshifts, see A.E. Hornschemeier et al., in preparation.

Among the ten OBXF galaxies at $z<0.15$ in the CDF-N with \hst\ 
imaging, we find that $\approx20$\% are likely off-nuclear black hole
candidates, exhibiting both variability and association
with optically bright knots along spiral arms.   We find another three sources within
this group that have properties consistent with ULX sources. Since we
have detected all $z\simlt0.15$ galaxies of at least moderate optical luminosity
(see \S \ref{trulynormal}), this gives some indication of the total fraction of
normal galaxies harboring ULX sources at $z\approx0.1$--0.2. This ULX
fraction is consistent with that measured in the local Universe \citep{Colbert2002}.
The ULX fraction measured here is only a lower limit; 
even \chandra's sub-arcsecond spatial resolution
cannot resolve sources within the central $\approx 1$--2~kpc of the nucleus
\citep[offsets of $\sim1$~kpc are not expected if the object is a supermassive
black hole, see discussion in ][]{Colbert2002}.  

\section{Conclusions and Future Prospects}
\label{OBXF_future}

	We have analyzed optically bright, 
X-ray faint sources arising in an area of high-exposure within 
the CDF-N 2~Ms survey.   
The 2~Ms data have doubled the number of detected OBXF sources over
the 1~Ms data presented in Paper~V and \cite{BargerCatalog2002}, for the
first time providing a sufficiently large sample for detailed study.
Here we summarize our main findings.

\begin{itemize}
 
\item{We present
25 optical spectra, including five new, previously unpublished redshifts.
The OBXF population is found to be fairly diverse, but overall 
dominated by non-AGN sources.  Within the
OBXF population we find that roughly 14\% are Galactic stars. 
The remaining 86\% of the OBXF population is consistent with the 
X-ray emission from both quiescent/``normal" galaxies and
more X-ray active starbursts and LLAGN. We note that the 
X-ray emission from galaxies may include lower-level accretion onto
supermassive black holes.  For instance, the soft X-ray spectral nature of these
objects is also consistent with LLAGN 
\citep[e.g., X-ray measurements of M81's nucleus and other
lower-luminosity AGN in local neighborhood, ][]{Petre93,Colbert99}.}

\item{We find that the X-ray number counts of normal and starburst galaxies
is fairly steep (slope $\approx-1.7$) down to the CDF-N survey limits, 
in contrast to the flattening  of the number counts for the general X-ray 
source population.  }

\item{Many of the galaxies in the OBXF sample are on average more X-ray luminous 
than truly ``normal" galaxies, indicating that even with 2~Ms, we 
are still detecting starburst and lower-luminosity AGN in greater numbers
than truly quiescent galaxies.  We detect most normal galaxies
out to $z\approx0.15$ and some out to $z\approx0.3$.}

\item{The composite X-ray spectrum of the 36 extragalactic OBXF sources
is very soft ($\Gamma\approx2.0$).   The complete lack of hard X-ray emission from these
galaxies supports the picture that they are not a population of highly
obscured AGN.   We find that the OBXF galaxies 
contribute $\approx1$\% to the soft XRB. }

\item{There are several instances of off-nuclear ULX sources within the
OBXF population.  Our two most likely candidates exhibit evidence
for variability.  Both are observed at look-back times of $\approx 1$~billion
years, representing a new epoch of study for these objects.  
}

\end{itemize}
 
We note that this work, particularly the search for ULX sources at
large look-back times, will be improved greatly in the coming year 
with the very deep multi-color imaging of the \hst\ Advanced Camera for
Surveys (ACS) Great Observatories Origins Deep Survey (GOODS, P.I. Mauro Giavalisco).  
The ACS GOODS survey has covered nearly all of the high-exposure area (HEA) studied here, 
providing excellent morphological and other information.

\acknowledgments

We gratefully acknowledge the financial support of
NASA grant NAS~8-38252 (GPG, PI),
NASA GSRP grant NGT5-50247 (AEH),
NSF CAREER award AST-9983783 (WNB,DMA),
\chandra\ X-ray Center grant G02-3187A (AEH, WNB, FEB),
NSF grant AST99-00703~(DPS) and $Chandra$ fellowship grant PF2-30021 (AEH).
We thank R. Ciardullo for helpful discussions and H. Ebeling for providing a
thorough and constructive referee report.
We thank the entire \chandra\ team and the community for making
this rich data set available.  
This work uses data from the DEEP project, which was
 obtained with support of the National 
Science Foundation grants AST 95-29028 and AST 00-71198 
awarded to S. M. Faber.
We gratefully acknowledge all the creators and operators of the
W. M. Keck Observatory.
The Hobby-Eberly Telescope (HET) is a joint project of the University of Texas
at Austin,
the Pennsylvania State University,  Stanford University,
Ludwig-Maximillians-Universit\"at M\"unchen, and Georg-August-Universit\"at
G\"ottingen.  The HET is named in honor of its principal benefactors,
William P. Hobby and Robert E. Eberly.
The Marcario LRS is named for Mike Marcario of High Lonesome Optics who fabricated
several optics for the instrument but died before its completion.


\bibliographystyle{apj}
\bibliography{Hornschemeier}


\begin{deluxetable}{rrcrrrrrrrrrrrrcrrrrr}
\rotate
\tablecolumns{18}
\tabletypesize{\scriptsize}
\tablewidth{0pt}
\tablecaption {Optically Bright, X-ray Faint Objects in the CDF-N High Exposure Area \label{xraydata_OBXF}}
%
%
\tablehead{
\multicolumn{1}{c}{RA$^{\rm a}$}                                         &
\multicolumn{1}{c}{DEC$^{\rm a}$}                                        &
\multicolumn{1}{c}{Pos. err.$^{\rm b}$}				&
\multicolumn{3}{c}{Counts$^{\rm c}$}                     &
\colhead{$\Gamma^{\rm d}$}                                             &
\multicolumn{3}{c}{$f_{\rm X}$ ($10^{-16}$~erg~cm$^{-2}$s$^{-1}$)}      &
\multicolumn{1}{c}{$t_{\rm Eff}$ (Ms)$^{\rm e}$}			&
\colhead{$R$}                                                   &
\colhead{$\Delta R^{\rm f}$}                                                   &
\colhead{$R$}							&
\multicolumn{2}{c}{$\log{({{f_{\rm X}}\over{f_{\rm R}}})}$ }		
\\
\multicolumn{1}{c}{(J2000)}                                      &
\multicolumn{1}{c}{(J2000)}                       		&
\multicolumn{1}{c}{($^{\prime \prime}$) }						&
\multicolumn{1}{c}{FB}                       			&
\multicolumn{1}{c}{HB}                       			&
\multicolumn{1}{c}{SB}                       			&
\colhead{}							&
\colhead{FB}                                      		&
\colhead{HB}                                      		&
\colhead{SB}                                      		&
\colhead{FB}                                      		&
\colhead{}                                                      &
\colhead{($^{\prime \prime}$) }			&
\colhead{Src$^{\rm f}$}                                         &
\colhead{FB}                                                      &
\colhead{SB}                                                      
}
\startdata
12 35 55.43 & 62 15 05.0 & 0.8 & $ 83.1^{+ 15.1}_{- 14.0}$   & $<   24.4$                  & $ 71.0^{+ 11.8}_{- 10.7}$   & $>1.76$      &  4.66     & $<  3.19$ &  2.19     &  1.66 &18.58 & 1.4 & 2 & $-2.40$ & $-2.73$ \\  
12 35 56.25 & 62 16 17.4 & 0.8 & $199.9^{+ 19.6}_{- 18.4}$  & $<   24.4$                  & $189.5^{+ 16.7}_{- 15.7}$  & $>2.64$      &  7.74     & $<  2.82$ &  6.61     &  1.60 &17.96 & 0.4 & 2 & $-2.43$ & $-2.50$ \\  
12 35 59.72 & 62 15 50.0 & 0.7 & $ 60.8^{+ 13.8}_{- 12.7}$   & $<   22.6$                  & $ 50.4^{+ 10.8}_{-  9.8}$    & $>1.52$      &  3.84     & $<  3.09$ &  1.52     &  1.68 &18.80 & 0.1 & 2 & $-2.40$ & $-2.80$ \\  
12 36 02.17 & 62 15 49.2 & 0.7 & $<   26.5$                  & $<   20.9$                  & $  8.6^{+  7.7}_{-  6.5}$      & $ \sim 2.00$ & $<  1.42$ & $<  2.65$ &  0.27     &  1.62 &17.55 & 0.4 & 2 & $-3.33$ & $-4.06$ \\  
12 36 09.75 & 62 11 45.9 & 0.6 & $ 45.4^{+ 11.0}_{-  9.9}$    & $<   18.0$                  & $ 39.5^{+  8.8}_{-  7.7}$     & $>1.51$      &  2.87     & $<  2.43$ &  1.18     &  1.69 &18.25 & 0.9 & 2 & $-2.74$ & $-3.13$ \\  
\\ 
12 36 14.42 & 62 13 19.0 & 0.6 & $ 49.3^{+ 10.2}_{-  9.0}$    & $<   17.8$                  & $ 30.7^{+  7.7}_{-  6.5}$     & $>1.29$      &  3.31     & $<  2.39$ &  0.85     &  1.80 &20.60 & 0.3 & 1 & $-1.74$ & $-2.33$ \\  
12 36 16.81 & 62 14 36.0 & 0.6 & $ 16.4^{+  7.9}_{-  6.7}$     & $<   12.2$                  & $ 15.1^{+  6.4}_{-  5.3}$     & $ \sim 2.00$ &  0.79     & $<  1.41$ &  0.42     &  1.80 &21.14 & 1.2 & 1 & $-2.15$ & $-2.42$ \\  
12 36 19.45 & 62 12 52.4 & 0.6 & $ 33.6^{+  8.5}_{-  7.1}$     & $<   15.3$                  & $ 25.9^{+  7.8}_{-  6.2}$     & $ \sim 2.00$ &  1.81     & $<  1.95$ &  0.80     &  1.62 &20.68 & 0.2 & 1 & $-1.97$ & $-2.32$ \\  
12 36 22.53 & 62 15 45.2 & 0.6 & $ 43.2^{+ 10.0}_{-  8.8}$    & $ 18.8^{+  7.7}_{-  6.4}$     & $ 24.9^{+  7.5}_{-  6.3}$     & 1.06         &  3.16     &  2.55     &  0.65     &  1.88 &20.74 & 0.9 & 1 & $-1.70$ & $-2.39$ \\  
12 36 22.76 & 62 12 59.7 & 0.6 & $<   18.7$                  & $<   12.3$                  & $ 18.7^{+  6.3}_{-  5.2}$     & $ \sim 2.00$ & $<  0.93$ & $<  1.43$ &  0.53     &  1.77 &20.95 & 0.1 & 1 & $-2.15$ & $-2.39$ \\  
\\ 
12 36 23.00 & 62 13 46.9 & 0.6 & $ 34.7^{+  8.4}_{-  7.2}$     & $<   15.6$                  & $ 20.6^{+  6.4}_{-  5.3}$     & $ \sim 2.00$ &  2.05     & $<  2.19$ &  0.70     &  1.47 &20.89 & 0.6 & 1 & $-1.83$ & $-2.30$ \\  
12 36 25.39 & 62 14 04.8 & 0.3 & $518.7^{+ 25.4}_{- 24.3}$  & $ 33.4^{+  8.2}_{-  7.0}$     & $477.9^{+ 23.7}_{- 22.5}$  & 3.19         & 19.40     &  3.76     & 18.80     &  1.51 &17.32 & 0.3 & 1 & $-2.28$ & $-2.30$ \\  
12 36 31.66 & 62 09 07.3 & 0.6 & $ 40.9^{+ 10.9}_{-  9.7}$    & $<   19.7$                  & $ 34.5^{+  8.5}_{-  7.4}$     & $>1.31$      &  2.76     & $<  2.67$ &  0.98     &  1.77 &20.32 & 0.6 & 1 & $-1.93$ & $-2.38$ \\  
12 36 33.81 & 62 08 07.7 & 0.8 & $ 38.3^{+ 12.1}_{- 10.9}$   & $<   22.1$                  & $ 35.6^{+  9.3}_{-  8.1}$     & $>1.23$      &  2.68     & $<  3.03$ &  0.99     &  1.78 &15.48 & 0.5 & 2 & $-3.88$ & $-4.31$ \\  
12 36 37.18 & 62 11 35.0 & 0.6 & $ 21.1^{+  7.2}_{-  6.0}$     & $<   10.1$                  & $ 17.8^{+  6.0}_{-  4.8}$     & $ \sim 2.00$ &  1.07     & $<  1.21$ &  0.52     &  1.72 &18.90 & 2.2 & 1 & $-2.91$ & $-3.23$ \\  
\\ 
12 36 40.12 & 62 19 42.0 & 0.7 & $169.3^{+ 17.8}_{- 16.7}$  & $<   21.6$                  & $166.0^{+ 15.6}_{- 14.4}$  & $>2.63$      &  6.04     & $<  2.29$ &  5.31     &  1.74 &15.47 & 0.9 & 2 & $-3.53$ & $-3.59$ \\  
12 36 41.81 & 62 11 32.1 & 0.6 & $ 39.9^{+  8.8}_{-  7.6}$     & $<   12.0$                  & $ 33.6^{+  7.5}_{-  6.4}$     & $>1.73$      &  2.22     & $<  1.54$ &  1.02     &  1.69 &19.96 & 1.1 & 1 & $-2.17$ & $-2.51$ \\  
12 36 44.00 & 62 12 50.1 & 0.6 & $ 15.7^{+  6.4}_{-  5.2}$     & $<   11.4$                  & $ 13.6^{+  5.5}_{-  4.3}$     & $ \sim 2.00$ &  0.71     & $<  1.21$ &  0.35     &  1.93 &21.41 & 0.3 & 1 & $-2.09$ & $-2.39$ \\  
12 36 47.04 & 62 12 38.2 & 0.6 & $<   13.3$                  & $<    9.3$                   & $  9.2^{+  5.0}_{-  3.8}$      & $ \sim 2.00$ & $<  0.60$ & $<  0.98$ &  0.24     &  1.93 &20.98 & 1.4 & 1 & $-2.33$ & $-2.73$ \\  
12 36 48.37 & 62 14 26.4 & 0.6 & $ 36.0^{+  8.2}_{-  7.1}$     & $<   12.3$                  & $ 28.9^{+  7.0}_{-  5.9}$     & $ \sim 2.00$ &  1.61     & $<  1.31$ &  0.75     &  1.94 &18.81 & 0.5 & 1 & $-2.77$ & $-3.10$ \\  
\\ 
12 36 49.45 & 62 13 47.1 & 0.6 & $ 16.2^{+  6.2}_{-  5.0}$     & $<   11.4$                  & $ 14.0^{+  5.6}_{-  4.4}$     & $ \sim 2.00$ &  0.84     & $<  1.40$ &  0.42     &  1.67 &18.26 & 0.4 & 1 & $-3.27$ & $-3.57$ \\  
12 36 51.15 & 62 10 30.4 & 0.6 & $ 36.6^{+  8.7}_{-  7.6}$     & $<   13.6$                  & $ 33.0^{+  7.6}_{-  6.4}$     & $>1.61$      &  2.02     & $<  1.66$ &  0.92     &  1.83 &20.58 & 0.5 & 1 & $-1.96$ & $-2.31$ \\  
12 36 52.89 & 62 14 44.1 & 0.6 & $ 96.9^{+ 12.0}_{- 10.8}$   & $ 18.9^{+  6.7}_{-  5.4}$     & $ 79.6^{+ 10.5}_{-  9.4}$    & 2.10         &  4.34     &  2.19     &  2.44     &  1.73 &19.50 & 0.1 & 1 & $-2.06$ & $-2.31$ \\  
12 36 52.95 & 62 07 26.8 & 0.9 & $120.0^{+ 19.0}_{- 17.0}$  & $<   25.1$                  & $109.6^{+ 17.4}_{- 15.2}$  & $>2.12$      &  5.23     & $<  2.89$ &  3.30     &  1.76 &14.01 & 0.3 & 2 & $-4.18$ & $-4.38$ \\  
12 36 54.26 & 62 07 45.3 & 0.8 & $ 35.4^{+ 11.9}_{- 10.7}$   & $<   26.8$                  & $ 22.5^{+  8.4}_{-  7.3}$     & $ \sim 2.00$ &  1.74     & $<  3.15$ &  0.64     &  1.77 &19.70 & 0.2 & 2 & $-2.38$ & $-2.82$ \\  
\\ 
12 36 58.33 & 62 09 58.5 & 0.6 & $ 29.8^{+  8.7}_{-  7.4}$     & $<   17.3$                  & $ 25.3^{+  7.2}_{-  6.0}$     & $ \sim 2.00$ &  1.42     & $<  1.96$ &  0.69     &  1.82 &18.36 & 0.4 & 1 & $-3.00$ & $-3.31$ \\  
12 36 58.85 & 62 16 37.9 & 0.6 & $ 23.1^{+  7.9}_{-  6.7}$     & $<   11.9$                  & $ 18.4^{+  6.3}_{-  5.2}$     & $ \sim 2.00$ &  1.08     & $<  1.34$ &  0.50     &  1.85 &19.87 & 0.1 & 1 & $-2.52$ & $-2.85$ \\  
12 37 01.99 & 62 11 22.1 & 0.6 & $ 21.9^{+  7.4}_{-  6.2}$     & $<   11.7$                  & $ 19.3^{+  6.2}_{-  5.0}$     & $ \sim 2.00$ &  1.01     & $<  1.28$ &  0.51     &  1.89 &19.51 & 0.9 & 1 & $-2.69$ & $-2.99$ \\  
12 37 06.12 & 62 17 11.9 & 0.6 & $ 34.0^{+  9.6}_{-  8.3}$     & $ 19.0^{+  8.1}_{-  6.9}$     & $ 17.6^{+  6.8}_{-  5.7}$     & 0.74         &  2.98     &  2.80     &  0.46     &  1.87 &19.05 & 0.9 & 1 & $-2.41$ & $-3.22$ \\  
12 37 08.33 & 62 10 55.9 & 0.6 & $ 44.1^{+  9.6}_{-  8.3}$     & $<   16.8$                  & $ 34.4^{+  7.8}_{-  6.7}$     & $>1.46$      &  2.77     & $<  2.23$ &  0.98     &  1.76 &20.34 & 0.6 & 1 & $-1.92$ & $-2.37$ \\  
\\ 
12 37 15.94 & 62 11 58.3 & 0.6 & $ 20.3^{+  7.8}_{-  6.6}$     & $<   13.9$                  & $ 17.7^{+  6.3}_{-  5.2}$     & $ \sim 2.00$ &  1.06     & $<  1.72$ &  0.53     &  1.67 &18.43 & 0.7 & 1 & $-3.10$ & $-3.40$ \\  
12 37 16.82 & 62 10 07.9 & 0.6 & $ 18.1^{+  8.9}_{-  7.7}$     & $<   19.1$                  & $ 17.5^{+  7.0}_{-  5.8}$     & $ \sim 2.00$ &  0.97     & $<  2.45$ &  0.54     &  1.62 &20.71 & 0.4 & 1 & $-2.23$ & $-2.49$ \\  
12 37 18.51 & 62 08 12.3 & 0.9 & $<   34.2$                  & $<   26.5$                  & $ 29.1^{+  9.1}_{-  7.9}$     & $ \sim 2.00$ & $<  1.99$ & $<  3.71$ &  0.97     &  1.49 &20.14 & 2.0 & 2 & $-2.15$ & $-2.46$ \\  
12 37 21.60 & 62 12 46.8 & 0.6 & $ 29.8^{+  8.7}_{-  7.6}$     & $<   17.5$                  & $ 18.6^{+  6.5}_{-  5.4}$     & $ \sim 2.00$ &  1.81     & $<  2.52$ &  0.66     &  1.43 &18.88 & 2.3 & 1 & $-2.69$ & $-3.13$ \\  
12 37 23.45 & 62 10 47.9 & 0.7 & $<   27.7$                  & $<   21.3$                  & $ 20.9^{+  7.3}_{-  6.1}$     & $ \sim 2.00$ & $<  1.40$ & $<  2.58$ &  0.61     &  1.72 &19.58 & 3.0 & 1 & $-2.52$ & $-2.88$ \\  
\\ 
12 37 25.57 & 62 19 42.9 & 1.0 & $ 45.2^{+ 15.6}_{- 13.6}$   & $<   32.2$                  & $ 33.2^{+ 11.8}_{-  9.6}$    & $>0.82$      &  4.02     & $<  5.00$ &  0.92     &  1.76 &18.11 & 1.7 & 2 & $-2.65$ & $-3.29$ \\  
12 37 25.65 & 62 16 49.0 & 0.7 & $ 96.7^{+ 14.0}_{- 12.8}$   & $<   20.4$                  & $ 96.1^{+ 12.1}_{- 11.1}$   & $>2.19$      &  3.94     & $<  2.24$ &  2.82     &  1.82 &19.05 & 0.2 & 1 & $-2.28$ & $-2.43$ \\  
12 37 27.71 & 62 10 34.3 & 0.8 & $ 35.9^{+ 11.2}_{- 10.0}$   & $<   23.1$                  & $ 32.5^{+  8.6}_{-  7.5}$     & $>1.11$      &  2.88     & $<  3.49$ &  0.97     &  1.66 &17.85 & 1.7 & 2 & $-2.90$ & $-3.38$ \\  
12 37 30.60 & 62 09 43.1 & 0.9 & $ 67.2^{+ 13.2}_{- 12.1}$   & $<   31.1$                  & $ 34.5^{+  9.1}_{-  7.9}$     & $>0.90$      &  6.27     & $<  5.14$ &  1.05     &  1.61 &18.68 & 1.8 & 2 & $-2.23$ & $-3.01$ \\  
12 37 34.10 & 62 11 39.6 & 0.8 & $ 34.3^{+ 11.0}_{-  9.9}$    & $<   22.5$                  & $ 27.8^{+  8.3}_{-  7.2}$     & $ \sim 2.00$ &  1.79     & $<  2.79$ &  0.84     &  1.67 &19.27 & 0.5 & 2 & $-2.54$ & $-2.87$ \\  
\\ 
\\
\\
\\
\\
\\
\\
\\
\\
\\
\\
\\
\\
12 37 37.14 & 62 12 05.3 & 0.8 & $ 23.8^{+ 11.0}_{-  9.9}$    & $<   23.7$                  & $ 19.6^{+  8.1}_{-  7.0}$     & $ \sim 2.00$ &  1.27     & $<  3.02$ &  0.60     &  1.63 &19.20 & 0.9 & 2 & $-2.72$ & $-3.04$ \\  
12 37 37.99 & 62 16 31.3 & 0.9 & $ 52.4^{+ 14.6}_{- 12.4}$   & $<   23.7$                  & $ 49.2^{+ 12.8}_{- 10.6}$   & $>1.45$      &  3.46     & $<  3.32$ &  1.48     &  1.67 &13.20 & 0.7 & 2 & $-4.68$ & $-5.05$ \\  
12 37 42.22 & 62 15 18.5 & 0.9 & $189.3^{+ 18.8}_{- 17.7}$  & $ 78.2^{+ 14.0}_{- 12.8}$   & $108.7^{+ 13.4}_{- 12.3}$  & 1.10         & 14.70     & 11.40     &  3.09     &  1.73 &18.33 & 0.3 & 2 & $-2.00$ & $-2.68$ \\  
\enddata
\tablenotetext{a}{These are the X-ray source coordinates. The positional error
is $\approx1$\farcs0 for all the sources (see \S\ref{ACISobs_OBXF}).}
\tablenotetext{b}{Positional error for this source as determined in Paper~XIII.}
\tablenotetext{c}{Source counts are as measured in Paper~XIII.}
\tablenotetext{d}{This is the estimated photon index based on the X-ray band ratio.  Values of $\Gamma=2.00$
marked with a ``$\sim$" are assumed (see \S \ref{OBXF_summary}).
This value corresponds to the mean photon index of all the OBXF sources (see \S~\ref{stacked_spectrum_OBXF}).}
\tablenotetext{e}{The effective exposure time is as determined in Paper~XIII.}
\tablenotetext{f}{The $R$-band magnitudes are Vega-based and are calculated from the $V$ and $I$-band magnitudes 
 of (1) \cite{BargerHFF} and (2) the UH8K $V$ and $I$-band images described in \S \ref{optphotomOBXF}. The $\Delta R$ values are the distances determined between the X-ray source and the
optical source in arcseconds.}
\end{deluxetable}

\clearpage


\begin{deluxetable}{rrcrrcrrl}
\tablecolumns{9}
\tabletypesize{\footnotesize}
\tablewidth{0pt}
\tablecaption {Redshifts and Luminosities for OBXF Sources with Redshifts or Identifications \label{multiwav_OBXF}}
%
%
\tablehead{
\multicolumn{1}{c}{CXOHDFN Name$^{\rm a}$}                      &
\multicolumn{1}{c}{$\Gamma^{\rm b}$}					&
\multicolumn{1}{c}{$z^{\rm c}$}                     			&
\multicolumn{1}{c}{$\Delta_{z}^{\rm d}$}                     		&
\multicolumn{1}{c}{$z$}		  				&
\multicolumn{2}{c}{$\log{L_{\rm X}}^{\rm f}$}				&
\multicolumn{1}{c}{ID$^{\rm g}$}			  			
\\
\multicolumn{1}{c}{(J2000)}                                     &
\multicolumn{1}{c}{}                     			&
\multicolumn{1}{c}{}                     			&
\multicolumn{1}{c}{($^{\prime \prime}$)}			&
\multicolumn{1}{c}{Src$^{\rm e}$}              			&
\multicolumn{1}{c}{SB}                       			&
\multicolumn{1}{c}{FB}						
}
\startdata
123555.4+621505 & $>1.76$      & $  0.207 $ & 1.2 & 1 & 40.50     & 40.82     & $ {\cal   E } $  \\   
123556.2+621617 & $>2.64$      & $  0.295 $ & 0.4 & 1 & 41.32     & 41.39     & $ {\cal   C } $  \\   
123559.7+621550 & $>1.52$      & $  0.375 $ & 0.5 & 1 & 40.93     & 41.33     & $ {\cal   E } $  \\   
123602.1+621549 & $ \sim 2.00$ & $  0.086 $ & 0.4 & 2 & 38.75     & $<39.48$  & $ {\cal   A } $  \\   
123609.7+621145 & $>1.51$      & $  0.136 $ & 0.9 & 1 & 39.83     & 40.21     & $ {\cal   A } $  \\   
123614.4+621319 & $>1.29$      & $  0.454 $ & 0.7 & 1 & 40.87     & 41.46     & $ {\cal   A } $  \\   
123616.8+621436 & $ \sim 2.00$ & $  0.515 $ & 0.2 & 3 & 40.69     & 40.97     & $ {\cal   E } $  \\   
123619.4+621252 & $ \sim 2.00$ & $  0.474 $ & 0.2 & 2 & 40.89     & 41.24     & $ {\cal   E } $  \\   
123622.5+621545 & 1.06         & $  0.647 $ & 0.5 & 3 & 41.12     & 41.81     & $ {\cal   E } $  \\   
123622.7+621259 & $ \sim 2.00$ & $  0.472 $ & 0.4 & 2 & 40.71     & $<40.95$  & $ {\cal   A } $  \\   
123623.0+621346 & $ \sim 2.00$ & $  0.485 $ & 0.7 & 2 & 40.85     & 41.32     & $ {\cal   E } $  \\   
123625.3+621404 & 3.19         & $ -2.000 $ & 0.3 & 2 & \nodata   & \nodata   & $ {\cal   S } $  \\   
123631.6+620907 & $>1.31$      & $  0.845 $ & 0.5 & 1 & 41.58     & 42.03     & $ {\cal   A } $  \\   
123633.8+620807 & $>1.23$      & $ -2.000 $ & 0.4 & 2 & \nodata   & \nodata   & $ {\cal   S } $  \\   
123637.1+621135 & $ \sim 2.00$ & $  0.078 $ & 2.3 & 2 & 38.95     & 39.27     & $ {\cal   E } $  \\   
123640.1+621942 & $>2.63$      & $ -2.000 $ & 0.6 & 1 & \nodata   & \nodata   & $ {\cal   S } $  \\   
123641.8+621132 & $>1.73$      & $  0.089 $ & 1.4 & 2 & 39.37     & 39.71     & $ {\cal   E } $  \\   
123644.0+621250 & $ \sim 2.00$ & $  0.557 $ & 0.2 & 2 & 40.70     & 41.00     & $ {\cal   E } $  \\   
123647.0+621238 & $ \sim 2.00$ & $  0.321 $ & 1.3 & 2 & 39.96     & $<40.37$  & $ {\cal   E } $  \\   
123648.3+621426 & $ \sim 2.00$ & $  0.139 $ & 0.6 & 2 & 39.65     & 39.98     & $ {\cal   E } $  \\   
123649.4+621347 & $ \sim 2.00$ & $  0.089 $ & 0.3 & 2 & 38.99     & 39.29     & $ {\cal   A } $  \\   
123651.1+621030 & $>1.61$      & $  0.410 $ & 0.6 & 2 & 40.80     & 41.14     & $ {\cal   C } $  \\   
123652.8+621444 & 2.10         & $  0.322 $ & 0.3 & 2 & 40.98     & 41.23     & $ {\cal   A } $  \\   
123652.9+620726 & $>2.12$      & $ -2.000 $ & 0.6 & 2 & \nodata   & \nodata   & $ {\cal   S } $  \\   
123654.2+620745 & $ \sim 2.00$ & $  0.202 $ & 0.8 & 3 & 39.94     & 40.37     & $ {\cal   E } $  \\   
123658.3+620958 & $ \sim 2.00$ & $  0.137 $ & 0.5 & 2 & 39.60     & 39.91     & $ {\cal   E } $  \\   
123658.8+621637 & $ \sim 2.00$ & $  0.299 $ & 0.6 & 2 & 40.22     & 40.55     & $ {\cal   E } $  \\   
123701.9+621122 & $ \sim 2.00$ & $  0.136 $ & 0.8 & 2 & 39.46     & 39.76     & $ {\cal   E } $  \\   
123706.1+621711 & 0.74         & $  0.253 $ & 0.4 & 3 & 40.01     & 40.83     & $ {\cal   E } $  \\   
123708.3+621055 & $>1.46$      & $  0.423 $ & 0.7 & 2 & 40.86     & 41.31     & $ {\cal   C } $  \\   
123715.9+621158 & $ \sim 2.00$ & $  0.112 $ & 0.9 & 2 & 39.30     & 39.60     & $ {\cal   A } $  \\   
123716.8+621007 & $ \sim 2.00$ & $  0.411 $ & 0.5 & 2 & 40.57     & 40.83     & $ {\cal   E } $  \\   
123718.5+620812 & $ \sim 2.00$ & $  0.411 $ & 1.4 & 3 & 40.83     & $<41.14$  & $ {\cal   E } $  \\   
123721.6+621246 & $ \sim 2.00$ & $  0.106 $ & 2.7 & 2 & 39.34     & 39.78     & $ {\cal   A } $  \\   
123723.4+621047 & $ \sim 2.00$ & $  0.113 $ & 3.1 & 3 & 39.37     & $<39.73$  & $ {\cal   E } $  \\   
123725.5+621942 & $>0.82$      & $  0.277 $ & 1.2 & 1 & 40.40     & 41.05     & $ {\cal   E } $  \\   
123725.6+621649 & $>2.19$      & $ -2.000 $ & 0.7 & 1 & \nodata   & \nodata   & $ {\cal   S } $  \\   
123730.6+620943 & $>0.90$      & $  0.298 $ & 1.7 & 2 & 40.54     & 41.31     & $ {\cal   E } $  \\   
123734.1+621139 & $ \sim 2.00$ & $  0.202 $ & 0.7 & 2 & 40.05     & 40.39     & $ {\cal   E } $  \\   
123737.1+621205 & $ \sim 2.00$ & $  0.410 $ & 1.5 & 3 & 40.62     & 40.94     & $ {\cal   E } $  \\   
123737.9+621631 & $>1.45$      & $ -2.000 $ & 1.5 & 1 & \nodata   & \nodata   & $ {\cal   S } $  \\   
123742.2+621518 & 1.10         & $  0.069 $ & 0.1 & 1 & 39.62     & 40.30     & $ {\cal   E } $  \\   
\enddata
\tablenotetext{a}{IAU-registered names for CDF-N sources.}
\tablenotetext{b}{$\Gamma$ is as in 
Table~\ref{xraydata_OBXF}.}
\tablenotetext{c}{Spectroscopic redshift, $-2.000$ indicates spectroscopically identified stars} 
\tablenotetext{d}{Distance between the X-ray source and the
optical source with measured redshift. These offsets were determined
using the coordinates provided in that the relevant redshift catalog.
 $-1.0$ indicates that the slit was placed at the location of the 
X-ray source.  }
\tablenotetext{e}{Source for the redshift, ``1" is 
\cite{BargerCatalog2002}, ``2" is an identification made from an optical spectrum obtained as part of
our survey (see \S\ref{optspectra_OBXF}), and ``3" is from 
A. Barger et~al., in preparation.}
\tablenotetext{f}{Rest-frame luminosity or upper limits; units are $\log($~\lumin). } 
\tablenotetext{g}{Spectroscopic identification (see \S \ref{optspectra_OBXF}).}
\end{deluxetable}

\clearpage


\begin{deluxetable}{lrrrrrr}
\rotate
\tablecolumns{7}
\tabletypesize{\footnotesize}
\tablewidth{0pt}
\tablecaption {Morphological Classification of Extragalactic OBXF Sources in the HST Flanking-Fields Area \label{morph_table_OBXF}}
%
%
\tablehead{
\multicolumn{1}{c}{CXOHDFN Name (J2000)}                                &
\colhead{$z^{\rm a}$}                                           &
\colhead{Description$^{\rm b}$}					&	
\colhead{$C(I)^{\rm c}$}                                        &
\colhead{$A(I)^{\rm c}$}                                        &
\colhead{Error of $A(I)^{\rm c}$}                                  &
\colhead{Notes}                                                 
}
\startdata
J123637.1+621135          &   0.078 & spiral w/knots        &          2.80 &         0.199 &         0.018 &                               Possible off-nuclear \\
J123641.8+621132          &   0.089 & spiral                &          3.02 &         0.282 &         0.001 &                                        Off-nuclear \\
J123649.4+621347          &   0.089 & E/S0                  &          3.76 &         0.093 &         0.000 &                                            \nodata \\
J123721.6+621246          &   0.106 & Grand-design Sb       &          3.17 &         0.095 &         0.029 &                                        Off-nuclear \\
J123715.9+621158          &   0.112 & spiral                &          4.42 &         0.205 &         0.005 &                                            \nodata \\
J123723.4+621047          &   0.113 & disk/spiral           &          3.46 &         0.209 &         0.029 &               Possible off-nuclear, edge-on spiral \\
J123701.9+621122          &   0.136 & barred spiral         &          2.54 &         0.090 &         0.022 &                Possible off-nuclear, prominent bar \\
J123658.3+620958          &   0.137 & Sc spiral, tail       &          3.62 &         0.148 &         0.011 &                                            \nodata \\
J123648.3+621426          &   0.139 & spiral w/knots        &          3.10 &         0.455 &         0.000 &                                            \nodata \\
J123647.0+621238          &   0.321 & disk                  &          3.83 &         0.134 &         0.002 &                                            \nodata \\
J123652.8+621444          &   0.322 & E/S0                  &          4.20 &         0.043 &       \nodata &             near chip gap, parameters less certain \\
J123651.1+621030          &   0.410 & spiral                &          3.66 &         0.076 &       \nodata &             near chip gap, parameters less certain \\
J123708.3+621055          &   0.423 & edge-on disk          &          2.94 &         0.357 &         0.019 &                                            \nodata \\
J123614.4+621319          &   0.454 & E/S0                  &          3.73 &         0.122 &         0.010 &                                            \nodata \\
J123623.0+621346          &   0.485 & disk                  &          3.11 &         0.151 &         0.017 &                          Possible spiral structure \\
J123616.8+621436          &   0.515 & irregular             &          2.94 &         0.312 &         0.069 &                                    Possible merger \\
J123644.0+621250          &   0.557 & merger                &          3.07 &         0.315 &         0.002 &                      Another galaxy 3\farcs0 away. \\
J123622.5+621545          &   0.647 & spiral/tail           &          3.38 &         0.353 &         0.031 &                               Disturbed morphology \\
\enddata
\tablenotetext{a}{Sources for redshifts are listed in Table~\ref{multiwav_OBXF}. }
\tablenotetext{b}{These are the fairly broad morphological classifications that 
we have made based on the imaging data described in \S \ref{optphotomOBXF}.}
\tablenotetext{c}{Concentrations ($C$), asymmetries ($A$) and
asymmetry errors calculated using the algorithm of \cite{cc00}. 
These were calculated using the $HST$ f814w images, and the
$I$ indicates that this is close to a measurement in the $I$-band.
Concentration errors are difficult to compute, but for
galaxies at these redshifts, they are $\approx \pm 0.15$ in $C$.}
\end{deluxetable}

\clearpage


\begin{deluxetable}{cccc}
\tablecolumns{4}
\tablewidth{0pt}
\tablecaption {Breakdown of source types by 0.5--8 keV X-ray-to-optical flux ratio 
\label{sources_by_fxfR}}
%
%
\tablehead{
\colhead{}							&
\colhead{OBXF$^{\rm a}$}                                         &
\colhead{Intermediate$^{\rm a,b}$}                                &
\colhead{Full X-ray Sample (OBXF)}						
\\
\colhead{Object category }						&
\multicolumn{1}{c}{[$\log{({{f_{\rm X}}\over{f_{\rm R}}})} < -2$]}	&
\colhead{[$-1 < \log{({{f_{\rm X}}\over{f_{\rm R}}})} < -2$]}		&
\colhead{[All values of $\log{({{f_{\rm X}}\over{f_{\rm R}}})}$]}								
}
\startdata
Galactic Stars	&  $14^{+9}_{-6}$\%	&\nodata	&	$\approx3$\%   \\
``Normal"	&  $52^{+14}_{-11}$\%	&$14^{+19}_{-9}$\%   & $\approx12 $\%  \\
Starburst/LLAGN	&  $33^{+12}_{-9}$\% 	&$64^{+29}_{-21}$\%     &      $\approx 25$\% \\
AGN$^{\rm c}$		&  \nodata &	$21^{+21}_{-12}$\%      &	$\simgt 60$\%  \\
\enddata
\tablenotetext{a}{The percentages may not add to exactly 100\% 
due to roundoff error.}
\tablenotetext{b}{This column is based on results of Paper~XI. There were no X-ray detected
stars in that sample.}
\tablenotetext{c}{We have explicitly attempted to exclude luminous AGN from the
OBXF sample.  There may be lower-luminosity AGN in the sample; these are included
in the starburst/LLAGN percentage.}
\end{deluxetable}




\begin{deluxetable}{ccccccr}
\tablecolumns{7}
\tablewidth{0pt}
\tablecaption {Possible off-nuclear X-ray sources within the OBXF population \label{ULX_OBXF_candidates}}
%
%
\tablehead{
\colhead{CXOHDFN Name}                                          &
\colhead{$z$}                                         		&
\colhead{Offset}						&
\colhead{Offset}						&
\multicolumn{2}{c}{$\log{L_{\rm X}}$ }				&
\colhead{NOTES}	
\\
\colhead{Name}							&
\colhead{}							&
\colhead{($^{\prime \prime}$)}					&
\colhead{(kpc)}							&
\colhead{SB}							&
\colhead{FB}							&
}
\startdata
\hline
\multicolumn{7}{c}{Highly confident candidates} \\
\hline
123641.8+621132 & 0.089&  1.1  &  2.0 &  39.37 & 39.71 	 & Spiral galaxy in HDF-N, see Paper~II \\
123721.6+621246 & 0.106&  2.3  &  4.8 &  39.34 & 39.78	 & Grand-design spiral (Figure~\ref{123721pretty})\\ 
\hline
\multicolumn{7}{c}{Other candidates} \\
\hline
123637.1+621135 & 0.078&  2.2   &  3.5 & 38.95 &  39.27  & Clumpy spiral galaxy \\
123701.9+621122 & 0.136&  0.9   &  2.3 & 39.46 &  39.76  & Spiral with prominent bar \\
123723.4+621047 & 0.113&  3.0   &  6.6 & 39.37 & $<39.73$& Edge-on spiral \\
\enddata

\end{deluxetable}

\clearpage


\begin{flushleft}
\begin{figure}
\epsscale{1.00}
\figurenum{5b}
\centerline{\plotone{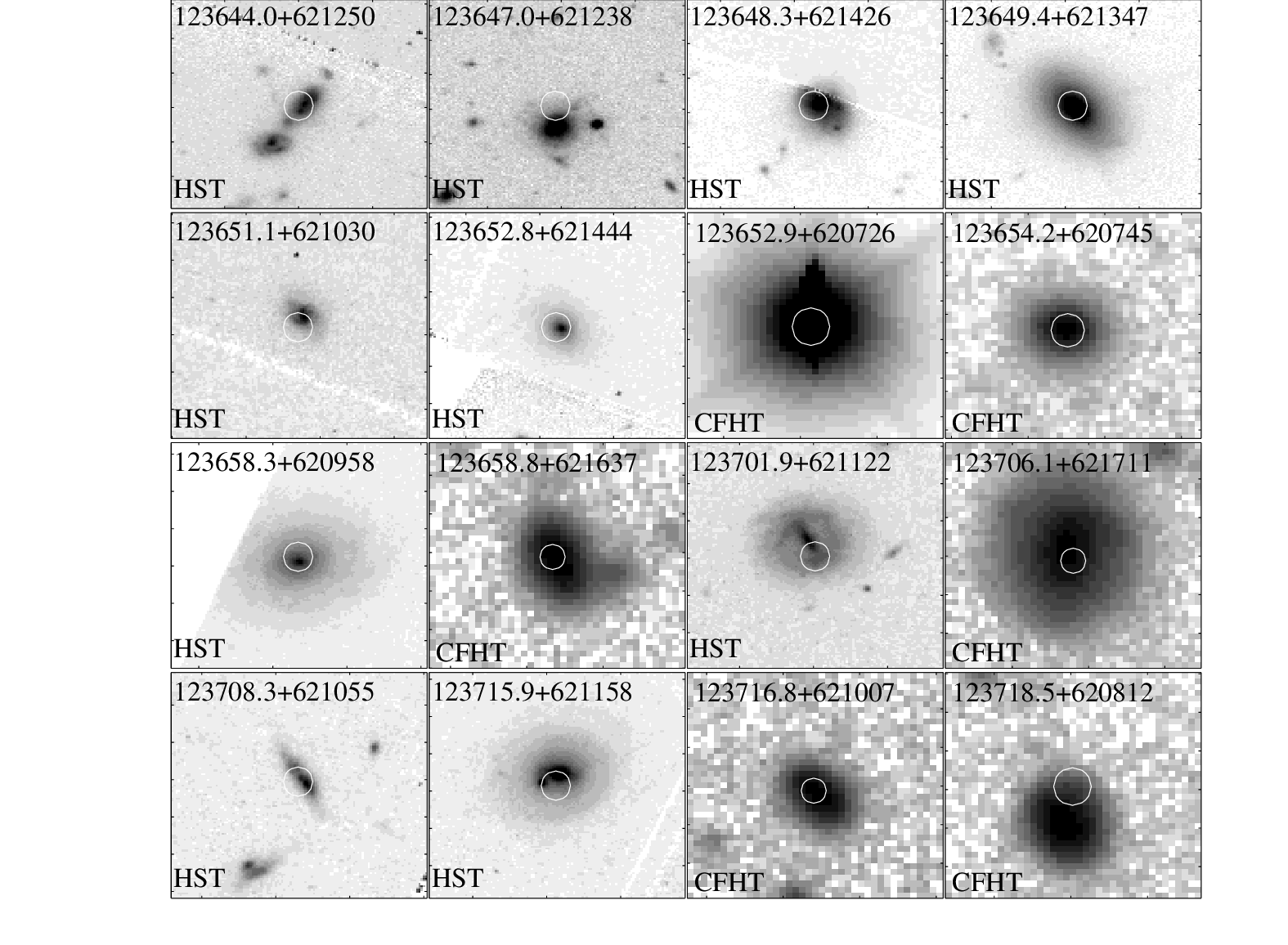}}
\caption[Optical cutout images, OBXF]{}
\end{figure}
\end{flushleft}

\begin{figure}
\epsscale{1.00}
\figurenum{5c}
\plotone{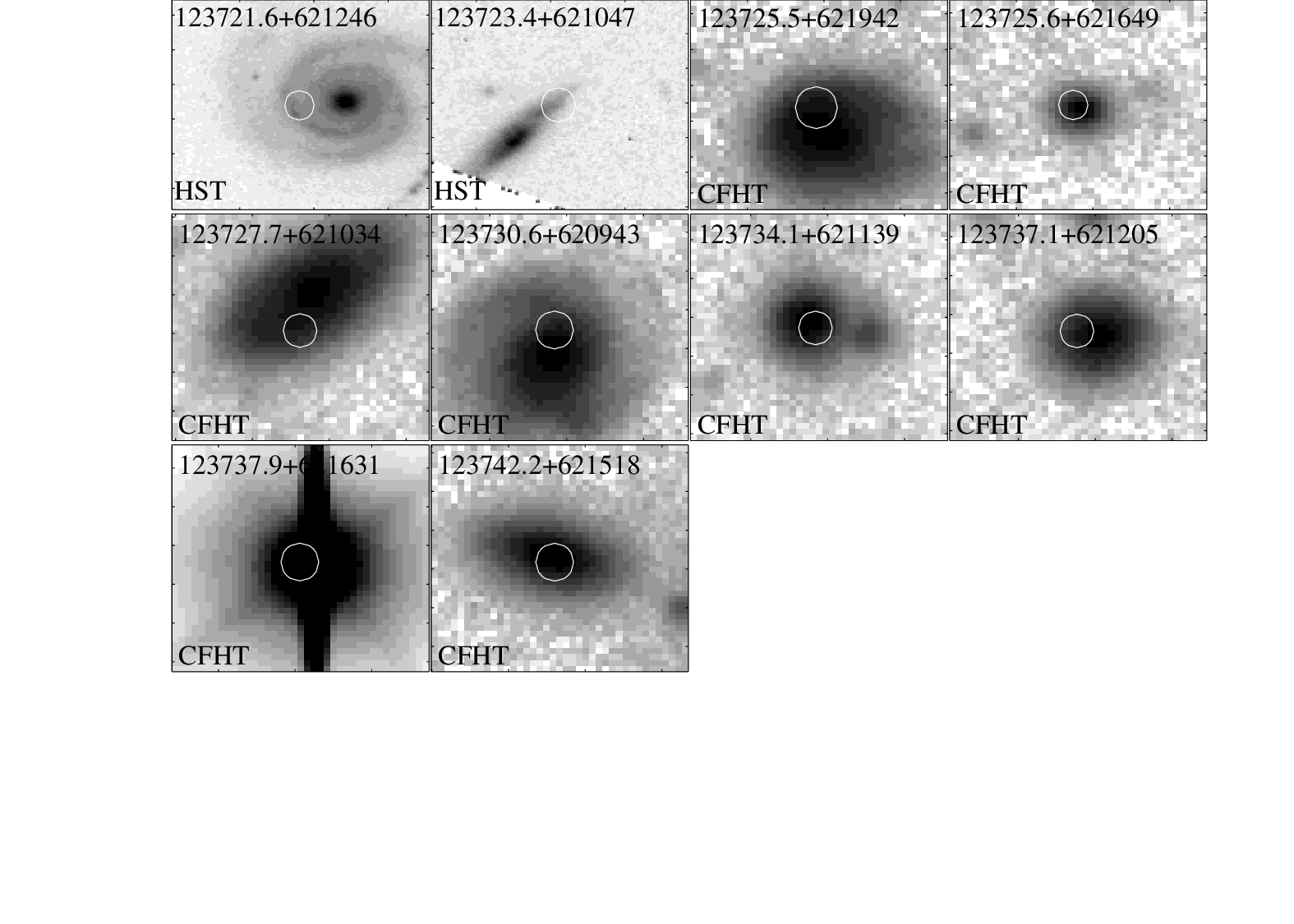}
\caption[Optical cutout images, OBXF]{}
\end{figure}


\begin{figure}
\figurenum{6}
\centerline{\includegraphics[scale=0.90,angle=0]{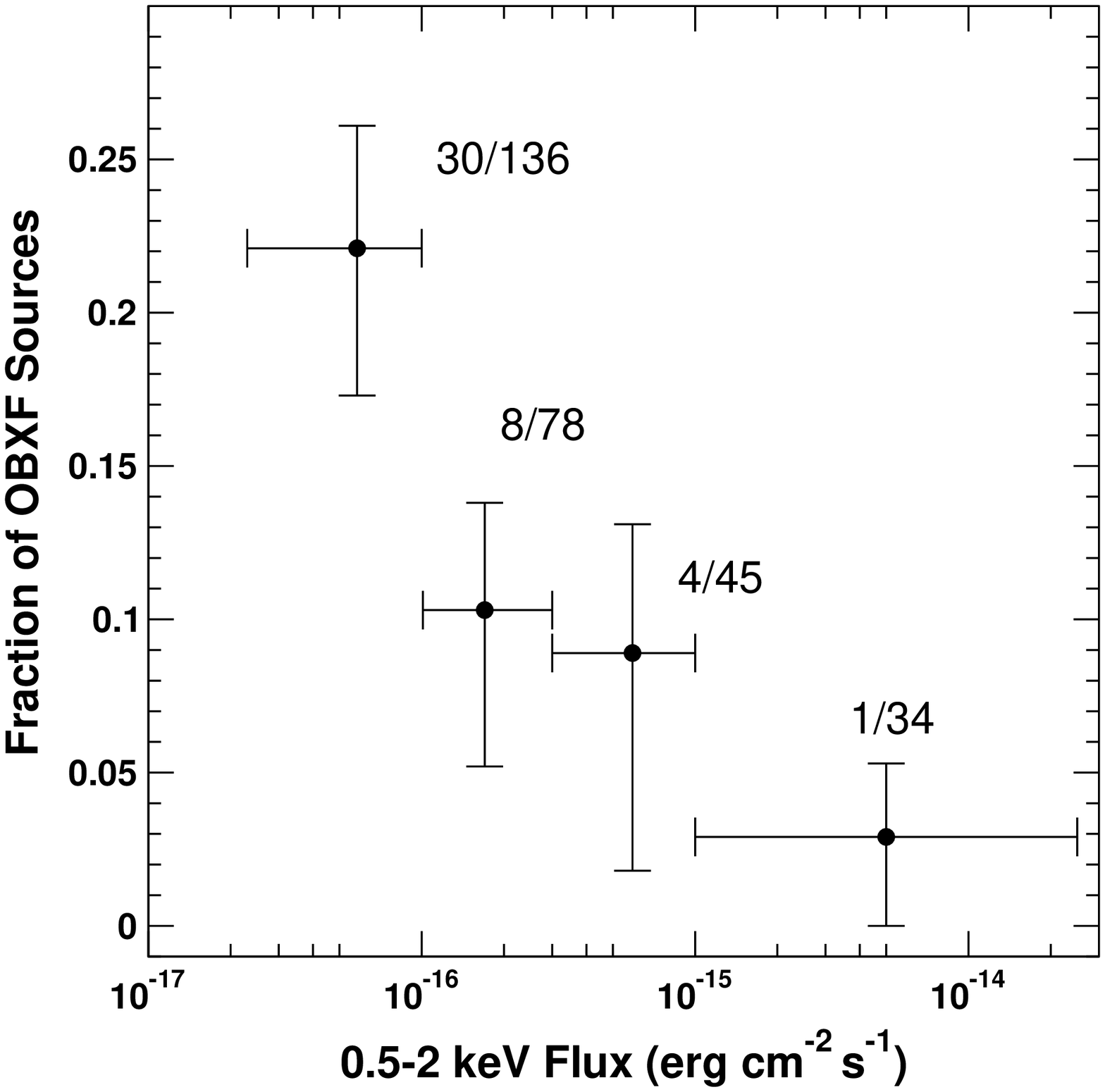}}
\caption[Fraction of OBXF Sources]{
Fraction of X-ray sources that are optically bright, X-ray faint
(OBXF) as a function of soft-band flux for the 2 Ms CDF-N survey HEA.   
The $x$-axis error bars 
indicate the range of X-ray fluxes 
over which the ratio was determined, and the $y$-axis error bars 
are the $1\sigma$ errors computed following
 \cite{Gehrels86}.  
\label{fraction_OBXF}}
\end{figure}


\begin{figure}
\epsscale{1.0}
\figurenum{7}

\plotone{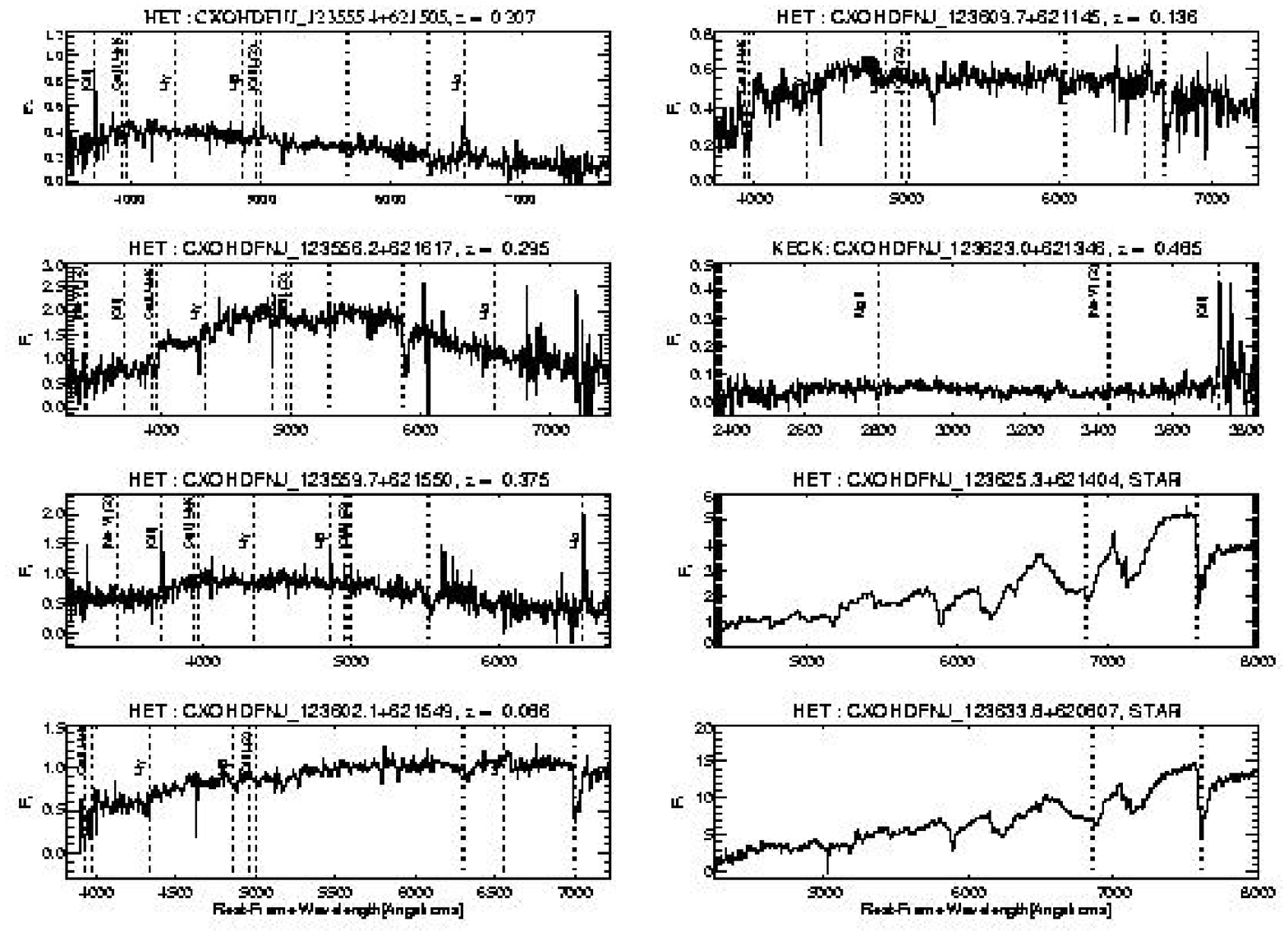}
\caption[Optical spectra of OBXF sample]{
Optical spectra of CDF-N OBXF sources.  For all spectra, the ordinate is 
relative F$_{\lambda}$ (erg~cm$^{-2}$ s$^{-1}$ \AA$^{-1}$).
No attempt has been made to place the spectra on an absolute 
spectrophotometric scale.  All spectra are plotted in rest-frame wavelength, and
for clarity we do not plot regions with extremely low signal-to-noise.
Several key optical transitions have been labeled.
A ``(2)" indicates that the next line redward 
is from the same element and ionization state.  The hash marked regions indicate
atmospheric absorption.
The spectral resolution is 
$\approx 14$--17~\AA\ for all the spectra.  Note that second-order
contamination is present in the HET spectra at wavelengths greater than
7700~\AA.  

\label{spectra1}}
\end{figure}


\begin{figure}
\epsscale{1.0}
\figurenum{7b}

\plotone{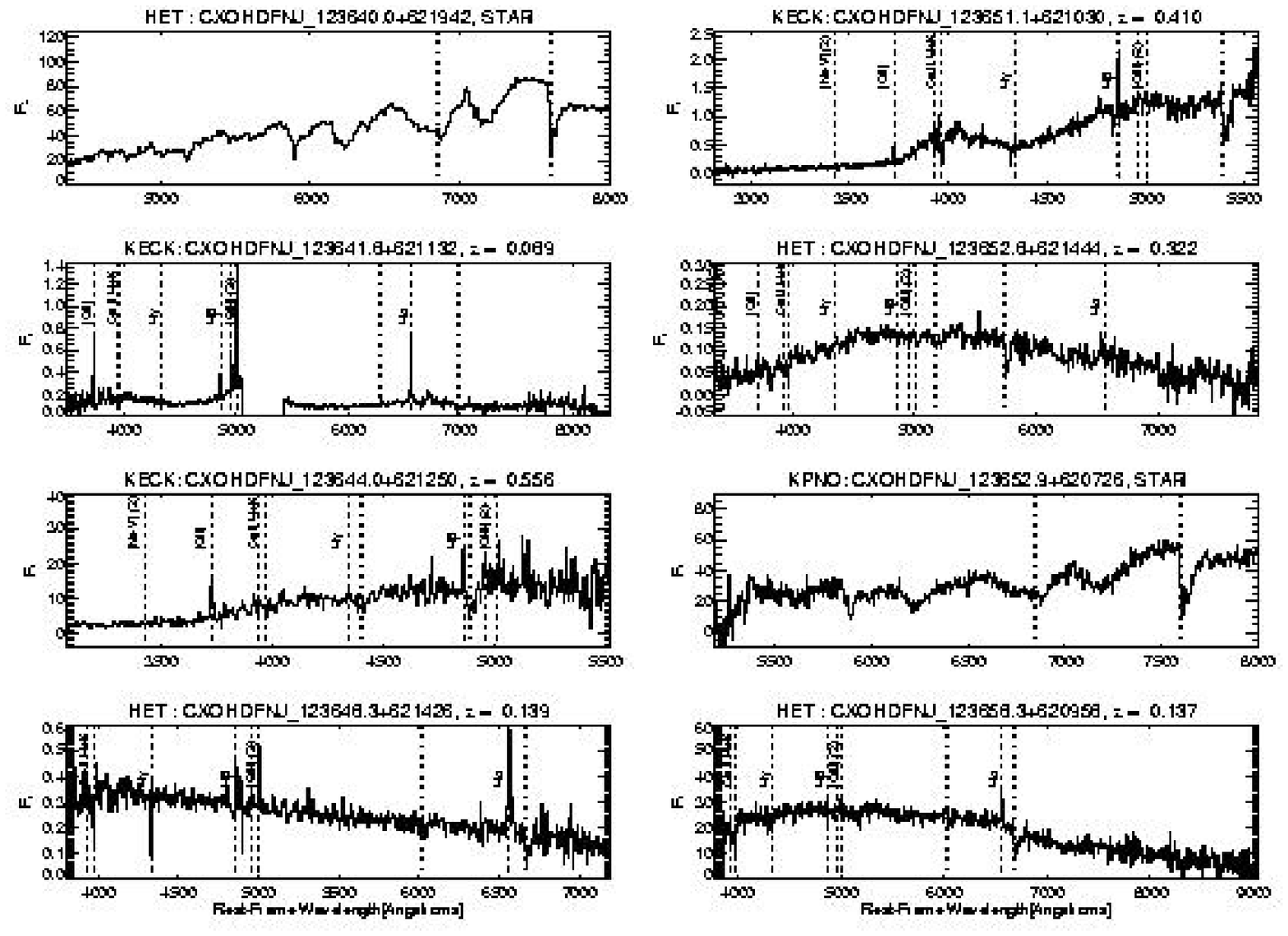}
\caption[Optical spectra of OBXF sample]{
\label{spectra2}}
\end{figure}

\begin{figure}
\epsscale{1.0}
\figurenum{7c}

\plotone{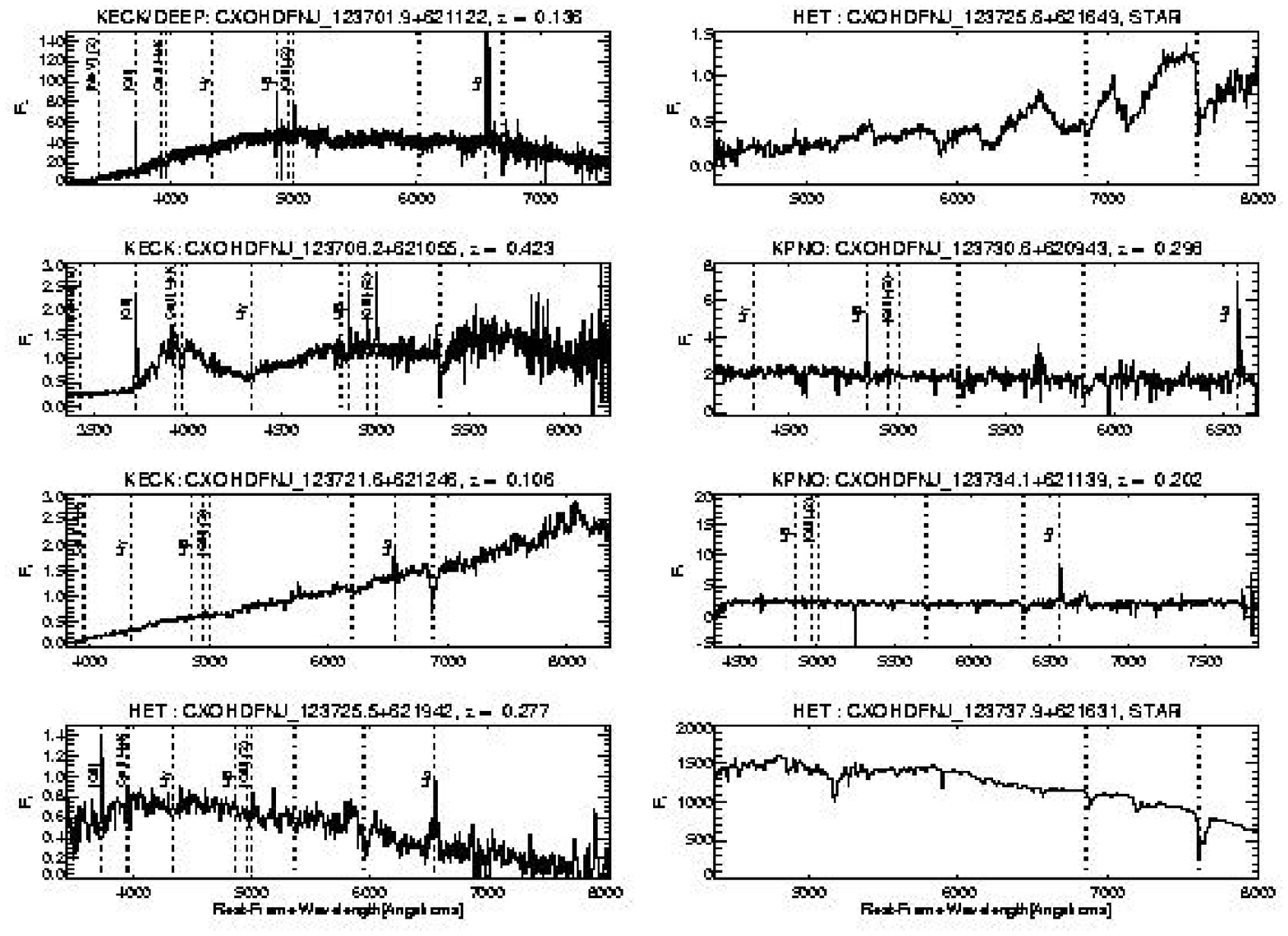}
\caption[Optical spectra of OBXF sample]{
\label{spectra3}}
\end{figure}
 

\begin{figure}
\epsscale{1.0}
\figurenum{7d}

\plotone{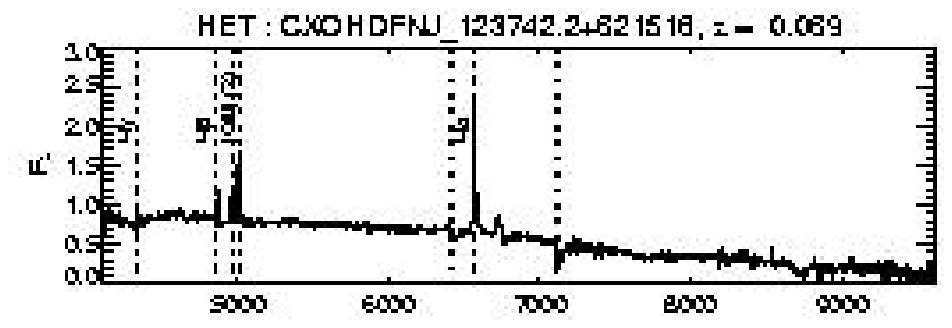}
\caption[Optical spectra of OBXF sample]{
\label{spectra4}}
\end{figure}


\begin{figure}
\epsscale{0.7}
\figurenum{8}
\plotone{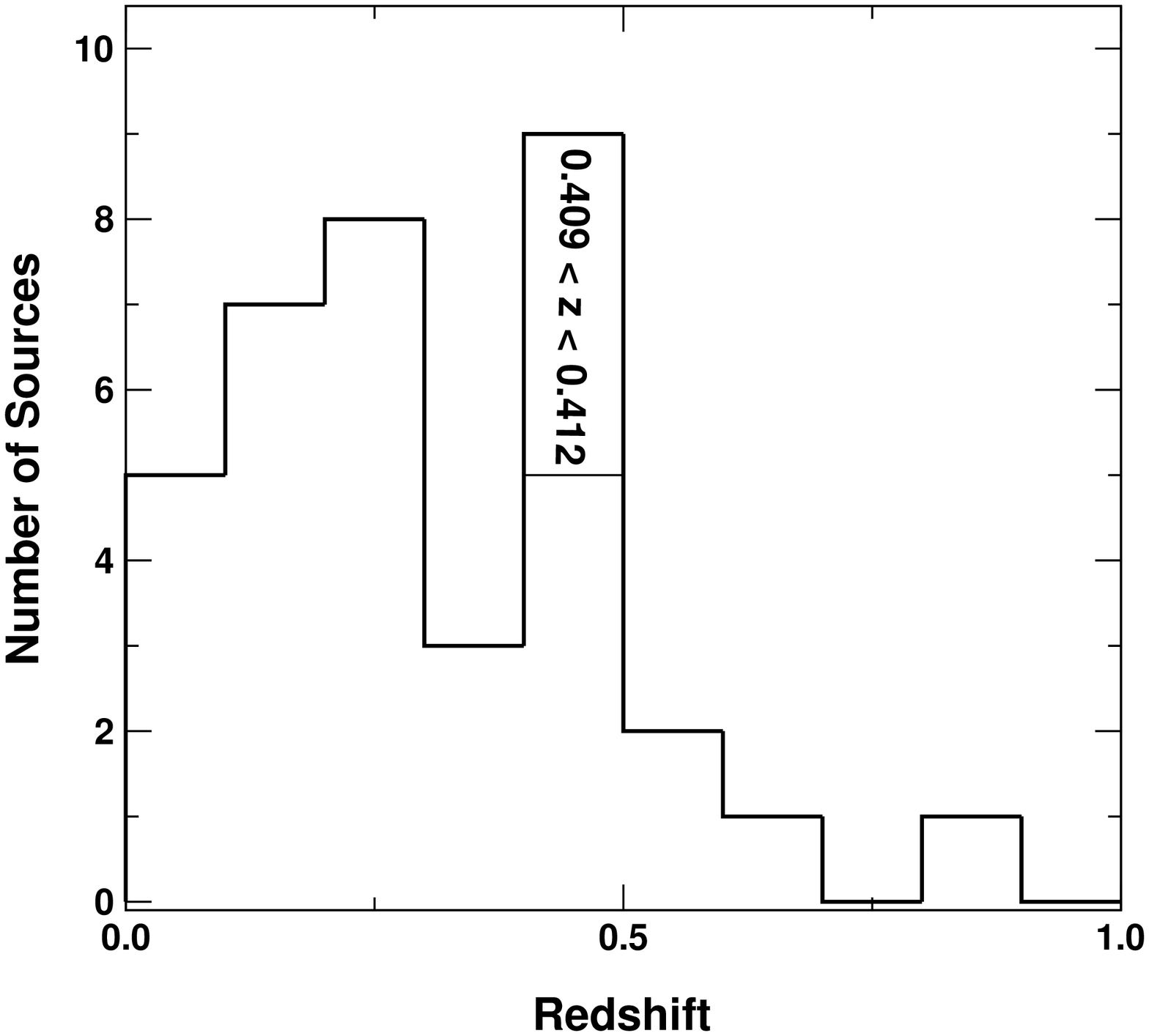}
\caption[Redshift distribution, OBXF]{Redshift distribution of the 40 identified extragalactic OBXF sources.  The marked bin indicates that four of the nine  
sources were detected in the very narrow redshift range $0.40 < z < 0.42$.
These correspond to two of the narrow redshift peaks discovered in the optical spectroscopic
survey of \cite{Cohen00} at $z=0.409$ and $z=0.421$.  It thus appears that deep
X-ray surveys trace similar structure, although the statistics are
limited \citep[see also][]{BargerCatalog2002}. 
\label{z_distribution}}
\end{figure}
\clearpage


\begin{figure}
\epsscale{0.7}
\figurenum{9}
\plotone{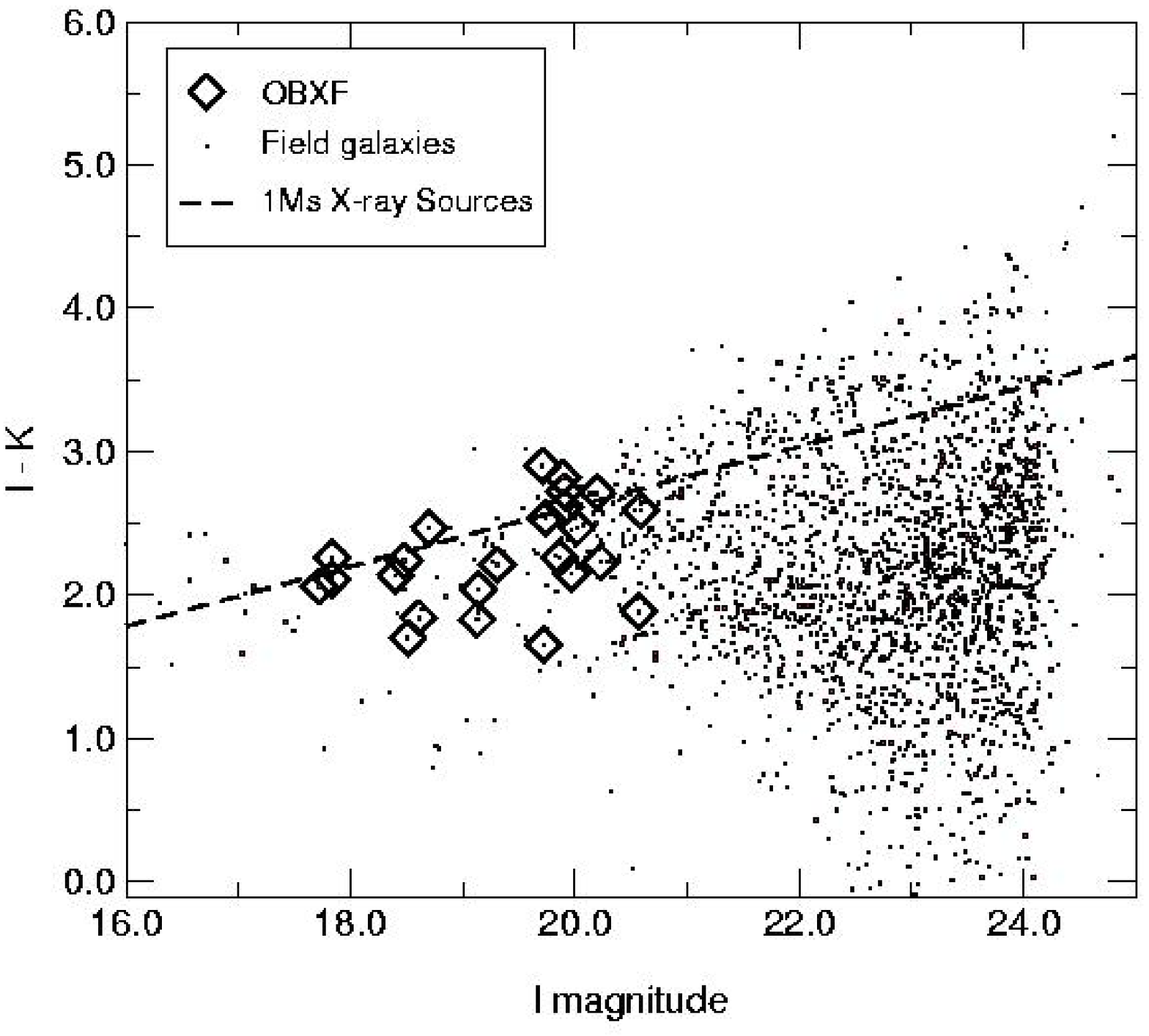}
\caption[Optical/near-infrared colors for the OBXF sample.]{
$I-K$ versus $I$ magnitude as calculated using the optical and
near-infrared photometry
of \cite{BargerHFF} for both field galaxies (dots) and OBXF 
sources (diamonds).  There are 27 galaxies measured in both the
$I$ and $K$ bands by \cite{BargerHFF}.
The line indicates the trend of redness found 
for the 1~Ms X-ray sources (Paper~VI).  Most (90\%)
of the 1~Ms X-ray sources have colors within $\Delta (I-K) = 0.6$ of
this line (Paper~VI).
\label{plot_colors_OBXF}}
\end{figure}


\begin{figure}
\figurenum{10}

\centerline{\includegraphics[scale=0.7,angle=0]{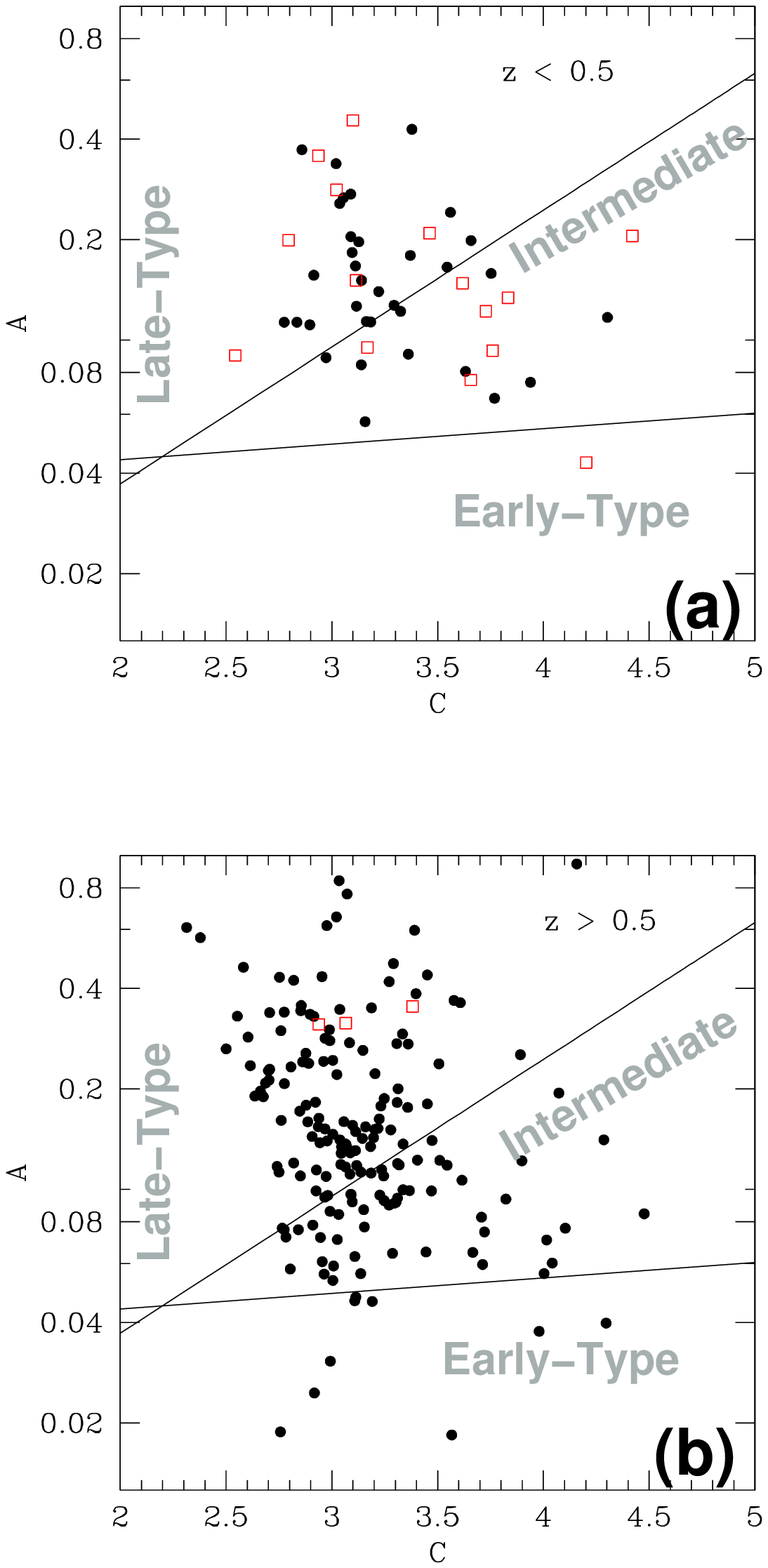}}
\caption[OBXF Morphology]{Asymmetry indices ($A$) versus 
concentration indices ($C$) for the OBXF galaxies (red boxes) and
the HDF-N field galaxy sample (black filled circles) calculated following
the quantitative measuring technique of \cite{cc00} and 
\cite{bershady00}.  (a) shows galaxies at $z<0.5$, and (b)
shows galaxies at $z>0.5$. The lines indicate
regions determined using the calibration of \cite{bershady00}.
\label{MORPH_OBXF}}
\end{figure}
\clearpage


\begin{figure}
\figurenum{11}

\centerline{\includegraphics[scale=0.95,angle=0]{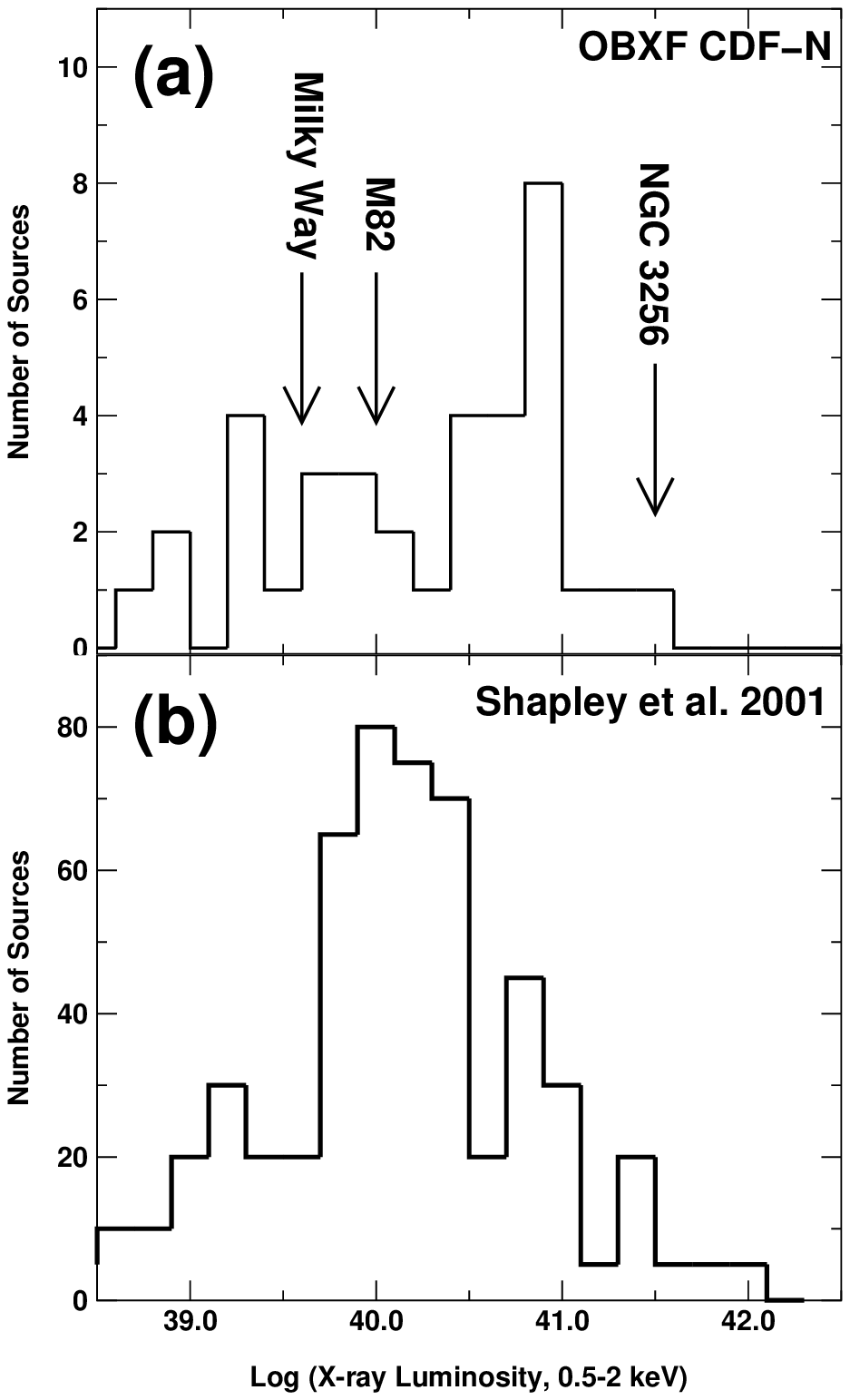}}

\caption[X-ray Luminosities of OBXF sources]{
 Histogram of 0.5--2.0~keV luminosity for (a) the 36 identified
extragalactic OBXF sources and (b) the late-type galaxies (S0, spiral, and irregular galaxies) 
from the \cite{Shapley01} \einstein sample.
We have converted the \cite{Shapley01} data for both
bandpass and cosmology.  Also plotted in panel (a) are the 0.5--2~keV X-ray luminosities
of the Milky Way~\citep{Warwick02}, M82~\citep{Griffiths00}, and NGC~3256~\citep*{Moran99}.
 
\label{Lx_histogram}}
\end{figure}
\clearpage


\begin{figure}
\figurenum{12}

\centerline{\includegraphics[scale=0.8,angle=0]{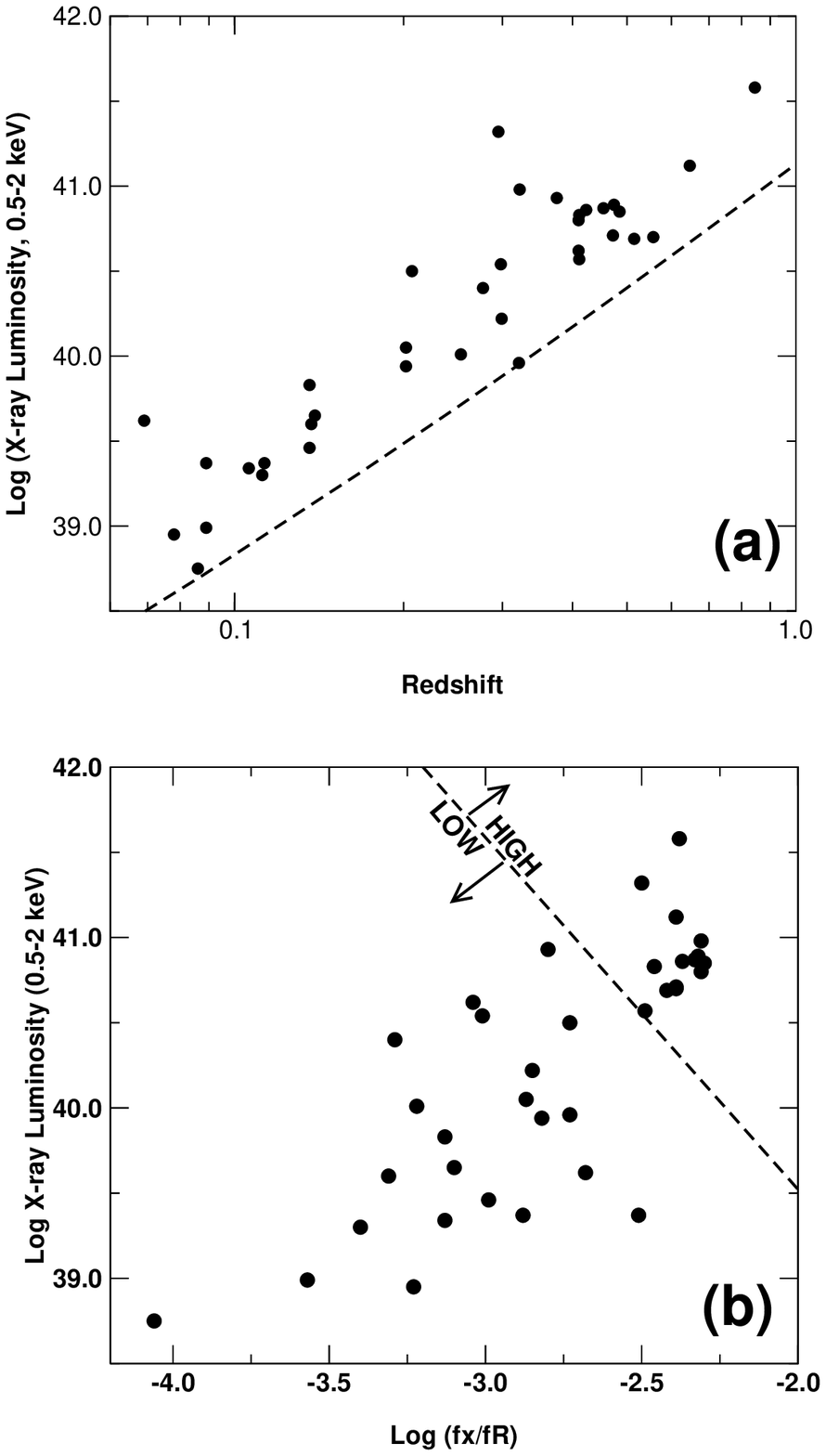}}

\caption[X-ray Luminosities of OBXF sources]{
(a) Luminosity-redshift diagram for the 36 extragalactic OBXF 
sources.  The dashed line marks the expected CDF-N sensitivity limit; this is
for $f_{\rm X} = 2.3\times10^{-17}$\flux (0.5--2.0~keV). 
(b) X-ray luminosity versus X-ray-to-optical flux ratio; the one
unidentified source is plotted at its X-ray-to-optical flux ratio
values with an open triangle.    The least 
X-ray luminous sources are also those with the least amount of X-ray
emission per unit optical emission.  The division between the ``high" and
the ``low" group is made for further comparison (e.g., X-ray spectral fitting
in \S \ref{stacked_spectrum_OBXF}).
\label{Lx_distribution}}
\end{figure}
\clearpage


\begin{figure}
\figurenum{13}
\centerline{\includegraphics[scale=0.60,angle=270]{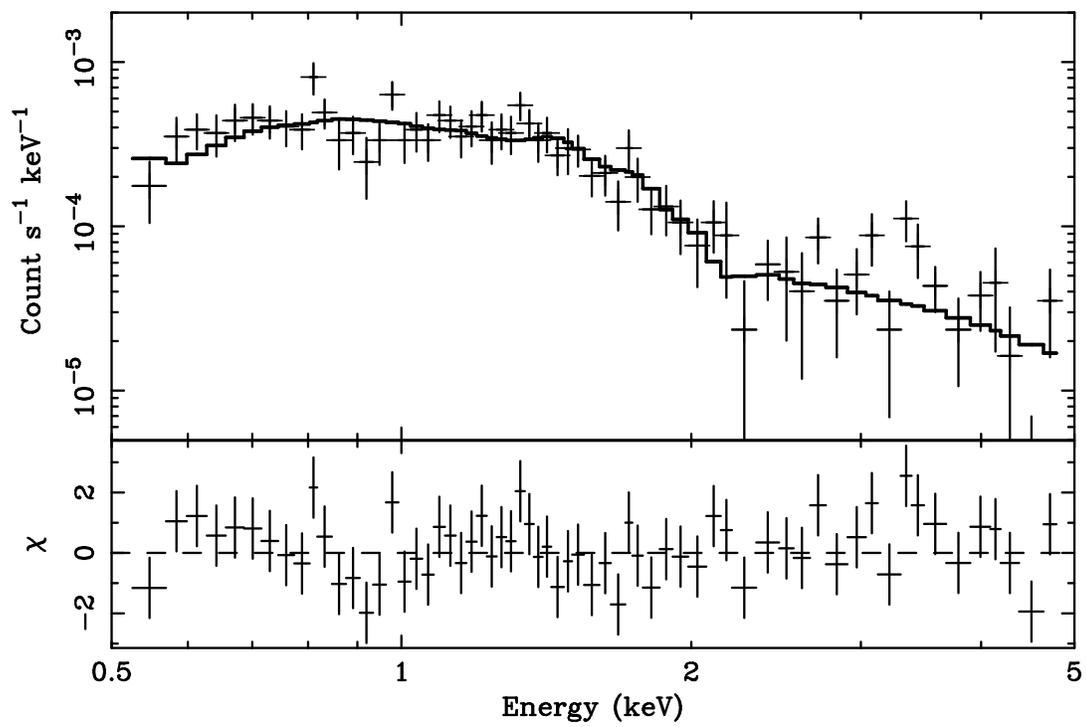}}

\caption[Stacked X-ray Spectrum for the Extragalactic OBXF Sources]{
Stacked, background-subtracted X-ray spectrum for the 36 identified
 extragalactic OBXF sources.
The spectrum is consistent with a power-law  model
($\Gamma=1.95$, see \S\ref{stacked_spectrum_OBXF}).
The Galactic column density is assumed. 
The lower panel shows the fit residuals in units of $\sigma$ with 
error bars of size unity.
\label{xray_spectrum_OBXF}}
\end{figure}
\clearpage


\begin{figure}
\epsscale{0.9}
\figurenum{14}
\centerline{\includegraphics[scale=0.6,angle=270]{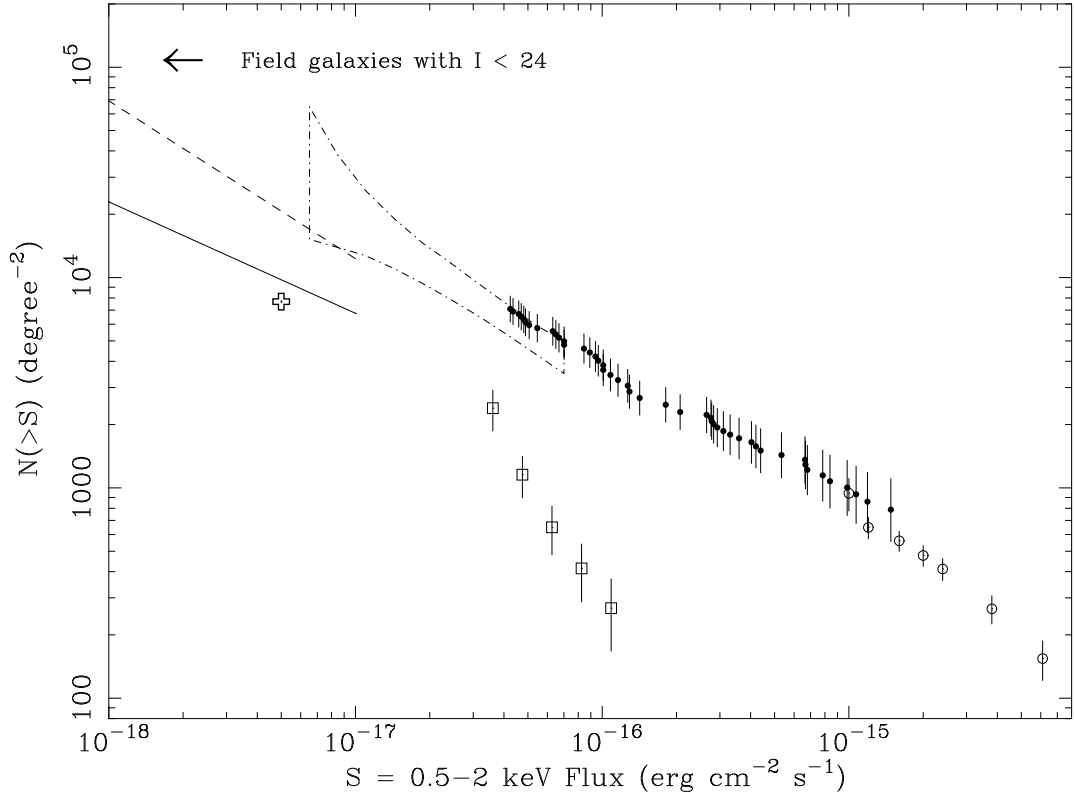}}

\caption[2 Ms Number Counts, OBXF]{
Number counts for the extragalactic OBXF population, marked with open boxes,
 as compared to other studies.  The CDF-N 1~Ms data (black filled circles) are from
Paper~V. The \rosat\ data are from \cite{Hasinger98}.
The dashed and solid black lines at faint X-ray fluxes show two
predictions of the galaxy number counts made by \cite{Ptak01}.
The dot-dashed lines mark the results of fluctuation analyses by \cite{Miyaji02}.
The cross marks the constraint from the 1~Ms stacking analysis
of Paper~VIII for relatively nearby spiral galaxies ($z\simlt1.4$).
The leftward-pointing arrow indicates the number density of field galaxies at $I=24$.
Note that the \cite{Miyaji02} analysis includes all X-ray sources at all
redshifts, including AGN and early-type galaxies, whereas
the Paper~VIII result only includes spiral
galaxies in the interval $0.4 < z < 1.5$.   The Paper~VIII
data point is thus lower as it represents a subset of
the normal galaxy population.
\label{logNlogS}}
\end{figure}
\clearpage



\clearpage

\appendix

{\bf CXOHDFN~J123622.5+621545}

This $z=0.647$ spiral galaxy has a high X-ray luminosity ($10^{41.8}$~\lumin,
0.5--8~keV) as compared to the rest of the OBXF population.  It also is
just marginally included in the OBXF sample as its full-band X-ray-to-optical
flux ratio is $-1.7$ (the soft-band X-ray-to-optical flux ratio is $-2.4$) and
its X-ray photon index is $\Gamma\approx1.1$.  The host galaxy is fairly
irregular, just narrowly below the asymmetry cutoff for mergers 
(see \S \ref{OBXF_morph}), so it is possible that this is a galaxy 
undergoing a vigorous starburst.  

{\bf CXOHDFN~J123706.1+621711}

This $z=0.253$ galaxy has a hard implied X-ray spectrum, $\Gamma\approx0.7$
but a low enough full-band X-ray-to-optical
flux ratio to be included in the sample ($-2.4$). 
The X-ray luminosity is $10^{40.8}$~\lumin (0.5--8~keV).
We do not have access to its optical spectrum to constrain its nature further.

{\bf CXOHDFN~J123742.2+621518} 

This source has a 0.5--8.0~keV luminosity of
$\approx10^{40.3}$~erg~s$^{-1}$ ($z=0.069$) and a fairly hard X-ray
spectrum ($\Gamma\approx1$).   
The X-ray source is coincident with the nucleus of its host
galaxy.  The optical spectrum of CXOHDFN~J123742.3+621518, obtained with the HET,
shows emission lines in \Hb, \OIII, [\OI] $\lambda 6300$, [\SII], \Ha, and \NII, so
we are able to perform some basic emission-line ratio diagnostics
\citep[e.g, ][]{Ho93, Dess00}. Unfortunately, [\OII]~$\lambda 3727$ lies at
a wavelength just short of the blue limit of our HET spectrum, so some
of the diagnostic line ratios, which require an estimate of [\OII] line
flux,  cannot be determined.  However, using the
line ratios of Figures~4, 5, and 6 of \cite{Ho93}, we are able to
ascertain that the optical spectrum is intermediate between HII
region/starburst-type spectra and Seyfert 2 type spectra. Both the
[\OI]/\Ha ~and \NII/\Ha ~ratios when compared with \OIII/\Hb
~indicate the object has a starburst-type spectrum.

\end{document}